\documentclass[acmsmall,screen]{acmart}

\usepackage{threeparttable}
\usepackage{multirow}
\usepackage{graphicx}
\usepackage{subfig}
\usepackage{colortbl}
\usepackage{afterpage}

\AtBeginDocument{%
  }

\setcopyright{acmcopyright}
\copyrightyear{2024}
\acmYear{2024}
\acmDOI{XXXXXXX.XXXXXXX}

\acmBooktitle{Proceedings of the 32nd ACM Symposium on the Foundations of Software Engineering (FSE '24), July 15--19, 2024, Porto de Galinhas, Brazil}

\acmPrice{15.00}
\acmISBN{978-1-4503-XXXX-X/18/06}

\begin{document}



\title[Accepted as a conference paper at FSE 2024]{LogSD: Detecting Anomalies from System Logs through Self-supervised Learning and Frequency-based Masking}

\author{Yongzheng Xie}
\email{yongzheng.xie@@adelaide.edu.au}
\affiliation{%
  \institution{The University of Adelaide}
  \city{Adelaide}
  \country{Australia}
}

\author{Hongyu Zhang}
\affiliation{%
  \institution{The University of Newcastle}
  \city{Newcastle}
  \country{Australia}}
\email{hongyu.zhang@newcastle.edu.au}

\author{Muhammad Ali Babar}
\authornote{corresponding author.}
\affiliation{%
  \institution{The University of Adelaide}
  \city{Adelaide}
  \country{Australia}
}
\email{ali.babar@adelaide.edu.au}

\renewcommand{\shortauthors}{Xie et al.}

\begin{abstract}
  Log analysis is one of the main techniques that engineers use for troubleshooting large-scale software systems. Over the years, many supervised, semi-supervised, and unsupervised log analysis methods have been proposed to detect system anomalies by analyzing system logs. Among these, semi-supervised methods have garnered increasing attention as they strike a balance between relaxed labeled data requirements and optimal detection performance, contrasting with their supervised and unsupervised counterparts. However, existing semi-supervised methods overlook the potential bias introduced by highly frequent log messages on the learned normal patterns, which leads to their less than satisfactory performance. In this study, we propose LogSD, a novel semi-supervised self-supervised learning approach. LogSD employs a dual-network architecture and incorporates a frequency-based masking scheme, a global-to-local reconstruction paradigm and three self-supervised learning tasks. These features enable LogSD to focus more on relatively infrequent log messages, thereby effectively learning less biased and more discriminative patterns from historical normal data. This emphasis ultimately leads to improved anomaly detection performance. Extensive experiments have been conducted on three commonly-used datasets and the results show that LogSD significantly outperforms eight state-of-the-art benchmark methods.
\end{abstract}





\maketitle

\section{Introduction}
\label{section:introduction}
  Modern software-intensive systems have grown notably in scale and complexity~\cite{meng2019loganomaly, yang2021semi, le2021log, xie2022loggd, zhang2022cat}. While these systems offer users a wealth of services, they also introduce new operational and maintenance challenges. Among these challenges is the task of detecting system issues and uncovering potential risks through the analysis of large amounts of system log data~\cite{zhang2019robust, meng2020semantic, le2021log, zhang2022cat}. Manually identifying anomalies based on massive log data has become an impractical endeavor~\cite{lin2016log, yang2021semi, xie2021logdp, xie2022loggd}. Therefore, automatically detecting system anomalies from logs has become essential.
    
  Over the past decades, a variety of log-based anomaly detection methods have been proposed~\cite{bodik2010fingerprinting,liang2007failure, chen2004failure, makanju2009clustering,lin2016log, du2017deeplog, zhang2019robust, meng2020semantic, zhang2020anomaly, xie2021logdp, yang2021semi, guo2021logbert,xie2022loggd, zhang2022cat}, which can be broadly classified into supervised, unsupervised, and semi-supervised categories. Supervised methods often achieve superior performance when a large amount of normal and abnormal data are available for training. However, their utility is limited by the labor-intensive requirements for data labeling and the scarcity of anomalous instances~\cite{yang2021semi, guo2021logbert, zhang2022cat}. Unsupervised methods eliminate the requirement for labeled data, including both normal and abnormal, but often yield sub-optimal performance due to the complete absence of label-based guidance~\cite{yang2021semi}. Semi-supervised methods offer a balanced solution, relaxing the need for scarce anomalous instances by leveraging more readily available normal data. This balance allows for a trade-off between stringent data requirements and optimal detection performance. In this paper, we focus on semi-supervised log anomaly detection.

  Semi-supervised anomaly detection methods typically start by training a model to capture normal patterns from historical data devoid of anomalies using a variety of techniques~\cite{wang2021multi, yang2021semi, farzad2020unsupervised, du2017deeplog, zhang2022cat, guo2021logbert}. They then identify anomalies based on deviations from these normal patterns learned by the model. However, the effectiveness of the learned patterns in these methods is often influenced by the occurrence frequency of the corresponding normal data in a dataset. In particular, the model tends to exhibit a stronger bias towards learning patterns associated with normal data that occurs more frequently. In the context of log-based anomaly detection, this phenomenon is common, which is reflected in the fact that some log events occur much more frequently than others. For example, in the BGL dataset~\cite{loghub}, the average occurrence ratio of frequent events to infrequent events is 120:1~\footnote{Further details on the meaning and setting of this frequency threshold are discussed in Section \ref{section: evaluation}.}. In the Spirit dataset~\cite{usenix}, this average ratio is 32:1. In such cases, frequently occurring patterns receive undue emphasis, while those that appear less often may be neglected or downplayed. Consequently, anomaly detection carried out with these biased patterns may encounter difficulties in distinguishing between normal instances containing less frequent events and those anomalous ones~\cite{xu2023fascinating}. 
  
\begin{figure*}[ht]
    \centering
    \subfloat[Issue: Biased Normal Pattern.]{\label{fig:problem_a} \includegraphics[width=0.4\textwidth]{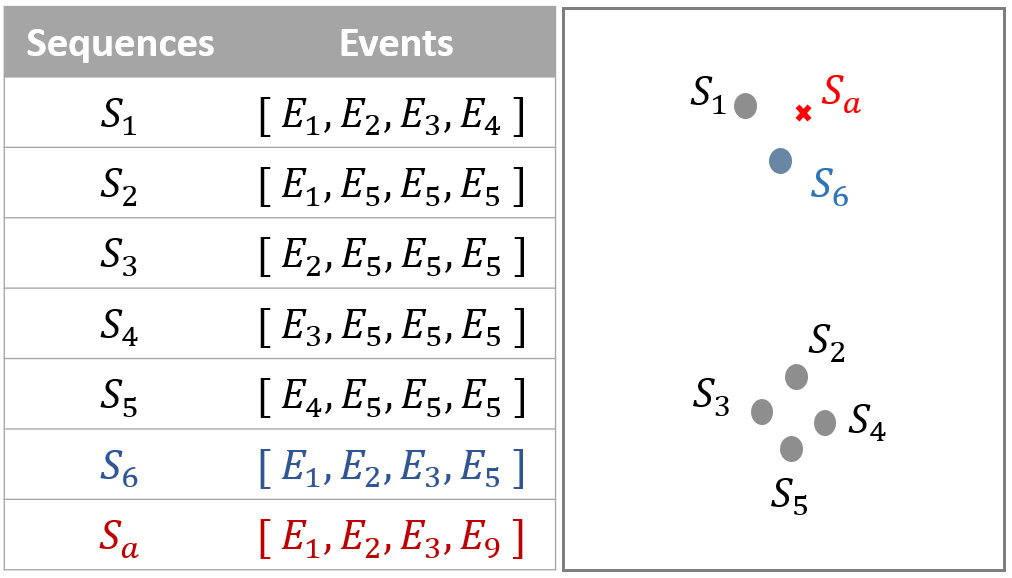}} 
    \subfloat[LogSD Solution.]{\label{fig:problem_b} \includegraphics[width=0.58\textwidth]{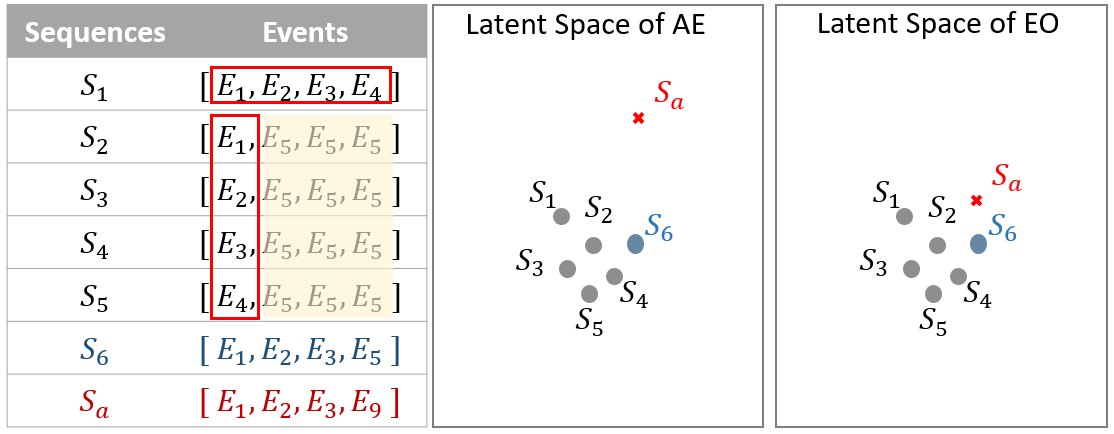}}  
    \caption{Illustrative example of the issue identified and the solution provided by our proposed method, LogSD. In both sub-figures, $S_1$ to $S5$ are normal training sequences. For testing, $S_6$ represents a normal sequence, and $S_a$ is anomalous. Sub-figure (a) demonstrates how frequent event $E_5$ in training sequences $S_2$ to $S_5$ can obscure the distinction between normal sequence and anomalous sequence in latent space; Sub-figure (b) depicts how LogSD addresses this issue by utilizing the contrast between two subnets: one to emphasize infrequent events and the other to focus on frequent events.}
    \label{fig:problem}
  \end{figure*}

  Figure \ref{fig:problem_a} gives an illustrative example of this issue. Consider a log dataset comprising six normal sequences, denoted as $S_1$ through $S_6$, along with an anomalous sequence $S_a$. Each sequence contains four events, which are listed in the 'Events' column. Sequences $S_1$ and $S_6$ contain four distinct events, while each of the sequences $S_2$ to $S_5$ comprises a unique event and three common events ($E_5$). The anomalous sequence $S_a$ shares three common events with $S_1$ ($E_1, E_2$ and $E_3$) and also includes an anomalous event ($E_9$). Let's consider that $S_1$ to $S_5$ are used for training, while $S_6$ and $S_a$ are designated for testing.
  During the training phase, when a model is used to learn representations of these sequences, the learned representations of sequences $S_2$ through $S_5$ stay closely in the latent space due to their common event $E_5$. However, the representation of $S_1$ diverges from others because of its dissimilarity to the sequences of $S_2$ to $S_5$. That is, the learned normal pattern, primarily represented by the representations cluster of $S_2$ to $S_5$, is dominated by the frequent occurrence of event $E_5$, while the normal pattern associated with $S_1$ is downplayed. During the testing phase, the representations of both $S_6$ and $S_a$ are similar and stay far from the learned normal pattern, largely due to their shared events with $S_1$. Their similarity introduces challenges in distinguishing the normal sequence $S_6$ from the anomalous sequence $S_a$.
    
  In this paper, we propose LogSD, a \textbf{Log}-based \textbf{S}elf-supervised learning method for anomaly \textbf{D}etection. LogSD addresses the problem of biased normal patterns by encouraging the model to focus on less frequent events during training. This strategy ensures that the representations of all normal sequences generated by two subnets, irrespective of the frequency of their events, are closely clustered in the respective latent space. Anomaly detection is then performed based on the contrast between these learned representations. Specifically, LogSD employs a dual-network architecture, consisting of an Auto-Encoder (AE) subnet and an Encoder-Only (EO) subnet. We also introduce a frequency-based masking scheme and a Global-to-Local (G2L) reconstruction paradigm to guide the model training focusing specifically on less frequent events. For an input sequence, two representations are generated: one from the AE subnet and another from the EO subnet. For a normal sequence, the two representations should be similar, whereas for an anomalous sequence, they will differ. The degree of difference between the two representations serves as an anomaly score to determine whether the input sequence is anomalous or not.

  To validate the efficacy of LogSD, extensive experiments have been conducted on three commonly used public log datasets. The results show that LogSD consistently outperforms eight state-of-the-art benchmark methods on the three datasets across varying window size settings. Specifically, LogSD improves the F1-score by at least 1.39\%, 12.7\%, and 82.6\% over the second-best method on the three respective datasets. Our ablation studies further confirm the importance of the key components in LogSD.

  In summary, our main contributions are as follows:
  \begin{enumerate}
   
    \item \textbf{A frequency-based masking scheme:} We propose an innovative masking scheme that leverages the occurrence frequency of log events in sequences to guide the masking of sequences. Our ablation study confirms that frequency-based masking can facilitate the model to achieve more stable and better performance in log-based anomaly detection.
    
    \item \textbf{A novel global-to-local reconstruction paradigm:} We introduce a novel global-to-local reconstruction paradigm that effectively incorporates overlooked information from existing self-supervised methods leveraging masked sequences as input. This paradigm enables LogSD to better capture the underlying normal pattern present in sequences.

    \item \textbf{A dual-network self-supervised anomaly detection approach:}  We propose an anomaly detection approach LogSD, which utilizes a dual-network architecture with three self-supervised learning tasks. One subnet is designed to emphasize infrequent events, while the other focuses on frequent events. By further integrating the frequency-based masking and global-to-local reconstruction techniques, LogSD achieves superior performance in log anomaly detection. To our best knowledge, this is the first dual-network self-supervised learning approach for log anomaly detection.
    
    \item \textbf{Extensive experiments:} We compare LogSD with eight SOTA baselines on three widely used public datasets. The results confirm the effectiveness of LogSD. In addition, extensive ablation experiments are conducted to show the efficacy of each major component in LogSD.
  \end{enumerate}

\section{Background}
\label{section:background}

\subsection{Log Data}
\begin{figure}[!ht] 
    \centering
    \includegraphics[width=0.8\linewidth]{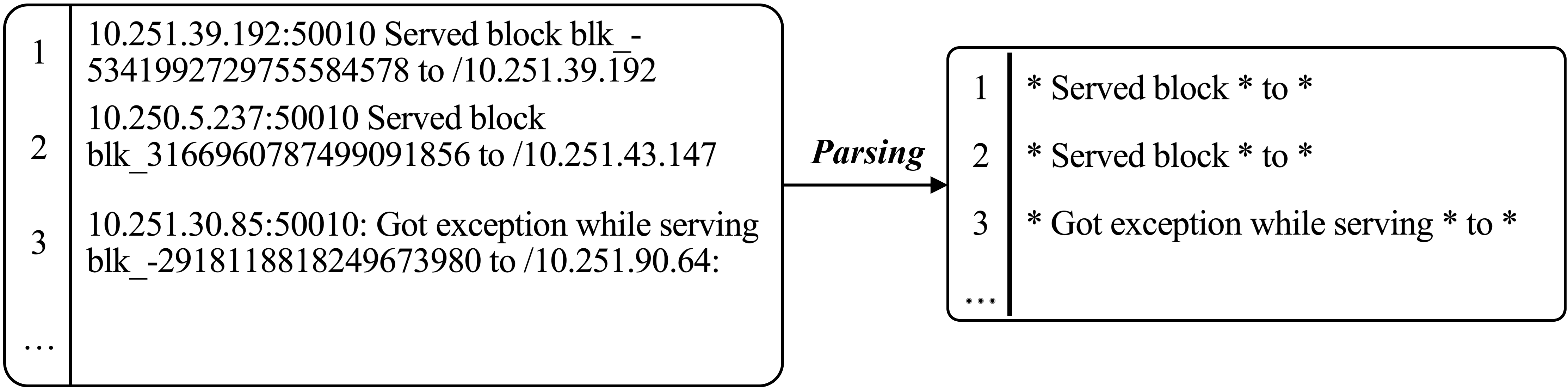}
    \caption{A snippet of HDFS (Hadoop Distributed File System) raw logs and events after parsing}
    \label{fig:LogsSnippet}
  \end{figure}

  A log message denotes a text sentence that is printed in the log file, which is used to record the run-time behaviors of systems~\cite{zhang2020anomaly, xie2021logdp, li2023they, fu2023empirical}. Log messages are usually semi-structured texts. Each log message consists of two parts: a constant part known as the log event or log template, and a variable part known as the log parameter. Log parser is a tool that is used to transform given log messages from semi-structured texts into structured log events and corresponding log parameters. Figure~\ref{fig:LogsSnippet} shows a snippet of raw logs and the results after they are parsed.

  An event sequence, also referred to as an instance, is parsed and grouped from a set of consecutive log messages that conform to chronological order. This succession of log events captures specific execution flows of the program that generate the log messages and is usually organized based on sessions or entries. Session-based log partition utilizes certain log identifiers to generate log sequences that encapsulate a series of interrelated operations. Alternatively, entry-based log partitioning often generate sequences by two strategies: fixed time window and fixed entry window. 

  \begin{itemize}
   \item The fixed time window approach entails the use of a consistent time window size, such as 15 minutes or 1 hour, for the creation of log sequences. Events within a sequence are timestamped within the predefined time frame. It is worth noting that when employing the fixed time window grouping strategy, the lengths of resulting sequences may vary in response to variations in system workloads~\cite{liao2020using}.
   \item The fixed entry window approach adopts a predetermined entry window size, such as 60 or 100 log messages, to produce log sequences. The sequences generated by this method maintain a consistent length. 
  \end{itemize}
  In addition, both fixed time window grouping and fixed entry window grouping have respective variants known as the time-sliding window and entry-sliding window strategies, which generate log sequences by overlapping two consecutive time or entry windows, respectively. For example, a 15-minute time window with a step size of 1, and an entry window of 100 logs with a step size of 1.

\subsection{Relevant Techniques}
\subsubsection{Self-supervised Learning}
\label{te:self_supervised_learning}
  Self-supervised learning (SSL) paradigm aims to harness intrinsic signals presented within the data itself for guiding model optimization~\cite{xie2022self}, culminating in the creation of representations tailored for diverse downstream tasks. This paradigm has achieved promising successes in natural language processing~\cite{devlin2018bert, ni2022best}, image processing~\cite{huang2022self}, and graph processing~\cite{tan2023s2gae}. Existing self-supervised learning approach can be roughly categorized into two groups: predictive methods and contrastive methods~\cite{xie2022self}. Predictive methods train deep learning models to predict certain labels obtained from the input data, such as masked input, statistical properties, and domain knowledge-based targets. In contrast, contrastive methods utilize pair-wise discrimination as their learning tasks. They apply transformations or augmentations to generate multiple views from given data samples and train models to distinguish between jointly sampled view pairs and independently sampled view pairs. 

\subsubsection{Knowledge Distillation}
\label{te:knowledge_distillation}
  Knowledge Distillation (KD) is a model training paradigm. It is originally designed to enhance a simplified model by transferring distilled insights from a more complex model. After the transferring, the simplified model is able to maintain a comparable level of accuracy to that complex mode. This paradigm is first introduced in the work~\cite{hinton2015distilling}, and has been extended to the field of anomaly detection~\cite{xiao2021unsupervised,salehi2021multiresolution, ma2022deep}. Knowledge distillation-based methods for anomaly detection typically involve two distinct sets of networks: the student network and the teacher network. The teacher network tends to have a higher capacity to learn knowledge from the dataset or additional data sources. These methods~\cite{xiao2021unsupervised,salehi2021multiresolution, ma2022deep} aim to train the student network to forecast the outputs of the teacher network. Anomalies are then identified when disparities emerge between the student network's outputs and those of the teacher network.

\subsubsection{One-class Classification}
  One-class classification technique is based on the assumption that a trained model can encapsulate all instances of normal data within a compact space, while anomalies inherently deviate from this norm. In the case of OC-SVM~\cite{scholkopf2001estimating}, normal samples are projected into a higher-dimensional space through a kernel function. This process establishes a hyperplane that optimally maximizes the separation between normal instances and anomalies. To enhance the aggregation of mapped data within the latent space, Deep SVDD~\cite{ruff2018deep} refines the neural network by minimizing the volume of a hyper-sphere encompassing the representations of the normal data. Consequently, this hyper-sphere serves as a boundary, effectively distinguishing anomalous instances from the normal ones. 

\section{Proposed Approach}
\subsection{Problem Definition}
  This proposed method, LogSD, aims to address the problem of end-to-end log-based anomaly detection at the sequence level in a semi-supervised manner. Specifically, given a set of normal sequences $\mathcal{X} = \{X_1, ..., X_n\}$, where $X_i$ denotes a normal sequence, we aim at learning an anomaly scoring function $f: \mathcal{X} \rightarrow \mathbb{R}$, parameterized by $ \theta $, such that for $\forall X^\prime_i, X^\prime_j \in \mathcal{X}^\prime, \mathcal{X} \bigcap \mathcal{X}^\prime \ne \emptyset$, $f(X^\prime_i; \theta) < f(X^\prime_j; \theta)$ if $X^\prime_i$ conforms to $\mathcal{X}$ better than $X^\prime_j$. Within the given sequences, each sequence can be denoted as $X_i = [E_1, ..., E_{|X_i|}]$, where ${E_k}$ indicates the $k$-th event within the sequence $X_i$ and $|X_i|$ represents the overall count of events within a session window or a predefined entry window. Moreover, every event can be further represented as $E_k = [token_1, ..., token_{|E_k|}]$, where $token_t$ corresponds to the $t$-th tokenized word, symbol, or numerical value within $E_k$, and $|E_k|$ signifies the total count of tokens encapsulated by $E_k$. To employ the underlying semantic relationships across $token_t \in V_{vocabulary}$, where $V_{vocabulary}$ denotes a token dictionary mapping each $token_t$ to a vector $x_i \in \mathbb{R}^m$, derived from some specific word embedding techniques utilized in Natural Language Processing (NLP) such as Word2Vec~\cite{nguyen2016integrating}, Glove~\cite{pennington2014glove}, FastText~\cite{joulin2016fasttext}, and BERT model~\cite{devlin2018bert}.

\afterpage{%
\begin{figure*}[ht]
  \centering
  \includegraphics[width=1.0\linewidth]{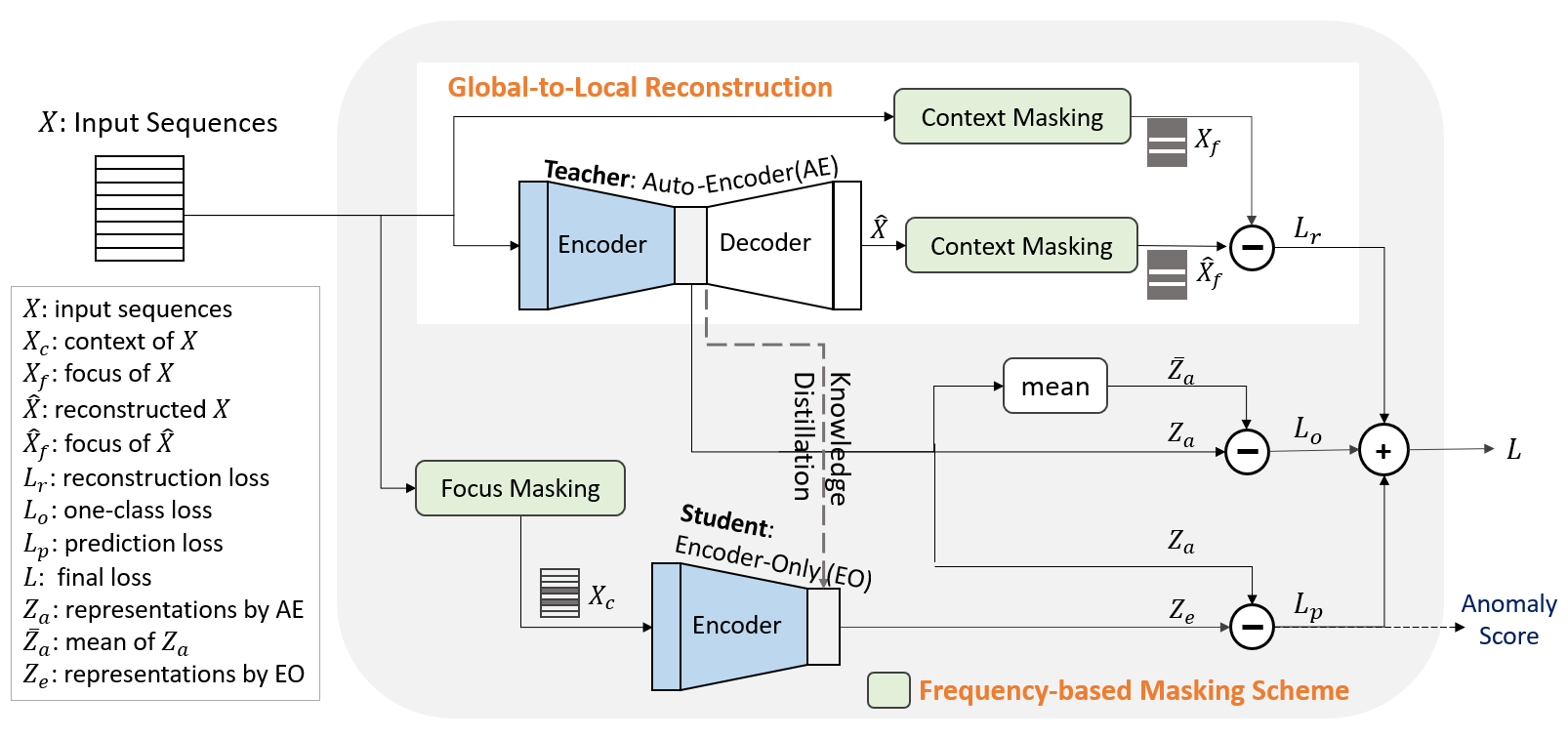}
  \caption{The overview of LogSD.
  }  
  \label{fig:framework_overview}
\end{figure*}
\vspace{-\baselineskip}
}

\subsection{Approach Overview}
\label{section:approach_overview}  
  As shown in Fig~\ref{fig:framework_overview}, LogSD consists of an auto-encoder subnet (AE) and an encoder-only (EO) subnet. The EO subnet has the same architecture as the encoder in the AE subnet. LogSD incorporates two key techniques to guide its training focus: frequency-based masking and a Global-to-Local (G2L) reconstruction paradigm. 

\subsubsection{Intuition of the Design}
  To clarify how LogSD works, let us apply it to the example depicted in Fig. \ref{fig:problem}. LogSD first categorizes events in the input sequences into frequent and infrequent events using a dynamically adjusted masking ratio, as elaborated in Section~\ref{subsection:frequency_based_masking_scheme}. Utilizing both the frequency-based masking and G2L reconstruction techniques, the AE subnet concentrates on reconstructing the infrequent events in the sequences, i.e., events $E_1$ to $E_4$ in this case, while ignoring the reconstruction of frequent event $E_5$. By doing so, the latent representations learned by AE emphasize the infrequent events present in input sequences.

  For the EO subnet, the training sequences are first modified by masking out the infrequent events, i.e., setting their values to zero. In this example, the values of sequences $S_2$ to $S_5$ become $[0, E_5, E_5, E_5]$ after masking, and are then used for training. The EO subnet is designed to learn the latent representations of the frequent events in an input sequence, when those infrequent events are missing in the input. During the training phase, the representations from the AE and EO subnets are optimized to be similar.

  During testing, for a normal sequence like $S_6$, its AE and EO representations are expected to be close to the normal pattern in the latent space, as depicted in the two plots in Fig. \ref{fig:problem_b}. Thus, its anomaly score, calculated as the difference between the two representations, would be low. On the other hand, for the anomalous sequence $S_a$, its representations generated from the two subnets will diverge. The EO representation is expected to be close to the normal pattern as the anomalous (also infrequent) event $E_9$ is set to zero during the masking process. While the AE representation diverges from the normal pattern because the AE subnet is unable to reconstruct the anomalous event $E_9$. This divergence results in a high anomaly score for $S_a$, making it easily detectable.

\subsubsection{Outline of the Design}
  In LogSD, frequency-based masking essentially categorizes events in the input sequences into two groups based on their occurrence frequencies: the context, containing frequent events, and the focus, containing infrequent events. For a given set of input sequences $X$, the AE subnet produces the reconstructions $\hat{X}$ and a set of representations $Z_a$ from its bottleneck layer. Frequency-based masking is then applied to both $X$ and $\hat{X}$, yielding the respective focuses, $X_f$ and $\hat{X}_f$, which are used to compute the reconstruction loss $L_r$. This loss guides the learning task of the AE subnet to better reconstruct $X_f$. Additionally, a one-class loss $L_o$ is computed as the difference between $Z_a$ and its mean $\overline{Z_a}$, encouraging the learned AE representations to cluster closely.

  For the EO subnet, frequency-based masking is applied to $X$ to obtain its context $X_c$, which is then fed into the EO subnet to yield a set of representations $Z_e$. A prediction loss $L_p$ is calculated as the difference between $Z_a$ and $Z_e$. This loss serves to ensure that the representations from the AE and EO subnets are similar for normal sequences.

  Finally, the three losses $L_r$, $L_o$, and $L_p$ are combined to form the final loss for training. During the inference phase, $L_p$ serves as the anomaly score, indicating the abnormality of a given sequence.

  Next, we elaborate on each major component of LogSD in the following subsections.

\subsection{Frequency-based Masking Scheme}
\label{subsection:frequency_based_masking_scheme}
  In this paper, we introduce a frequency-based masking scheme tailored specifically to the log analysis domain. In LogSD, event occurrence frequencies are calculated at the batch level. When the occurrence frequency of each event is available, the events with the lowest $\kappa$ occurrence frequency in the batch are selected to be the focus, while the remaining events constitute the context. The parameter $\kappa$ can be configured as a fixed value or as a ratio of unique events in a batch, serving as a hyper-parameter. Inspired by randomly masking scheme, the frequency-based masking can be further extended to a random frequency-based masking scheme. This entails dynamically generating a unique mask ratio in real-time for each batch. Specifically, the parameter $\kappa$ is sampled from a set $K = \{k_i\}_{i=1:N_k}$, where ${N_k}$ is the set cardinality. In this study, we empirically set $K = \{0.05, 0.1, 0.15, 0.2, 0.3\}$, where ${N_k}$ is 5. Subsequently, the randomly determined ratio $\kappa$ is utilized to facilitate the masking processing.

  Based on the grouped focus events and context events, two distinct masking processing are employed: focus masking and context masking. In focus masking, infrequent events, termed as \textit{focus}, are masked out from the input sequences, resulting in what we refer to as the \textit{context}. Conversely, in context masking, frequent events, or the \textit{context}, are masked out to produce the \textit{focus}. As depicted in Fig~\ref{fig:framework_overview}, in LogSD, context masking is performed prior to the computation of the reconstruction loss $L_r$. This ensures that the loss is calculated based on the focus, guiding the AE subnet to learn the reconstruction of infrequent events. In contrast, focus masking is applied to the sequences $X$ to produce the context $X_c$, which serves as the input to the EO subnet. This enables the EO subnet to focus on learning the representations of $X$ with the absence of input information about these events. By doing so, the frequency-based masking strategy ensures LogSD learn the less biased normal patterns in both AE and EO representations by emphasizing the learning of infrequent events.

\subsection{Global-to-Local Reconstruction Paradigm}
\label{subsection:G2L_reconstruction_paradigm}
\begin{figure*}[ht]
    \centering
    \subfloat[Global Reconstruction.]{\label{fig:global_reconstruction} \includegraphics[width=0.32\textwidth]{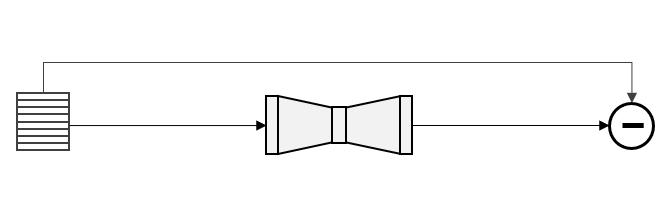}} 
    \subfloat[Local Reconstruction.]{\label{fig:local_reconstruction} \includegraphics[width=0.32\textwidth]{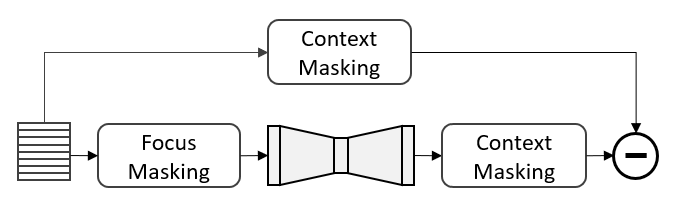}}
    \subfloat[G2L reconstruction.]{\label{fig:G2L_reconstruction} \includegraphics[width=0.32\textwidth]{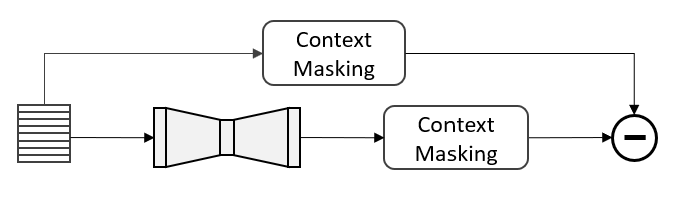}}
    \caption{Different Reconstruction Paradigms.}  
    \label{fig:reconstruction_paradigms}
  \end{figure*}

  The proposed Global-to-Local (G2L) reconstruction paradigm distinguishes itself from existing reconstruction approaches, including both global and local paradigms as described in Section~\ref{subsection:existing_reconstruction_paradigm}. As depicted in Fig.~\ref{fig:global_reconstruction}, global reconstruction employs the complete input sequence (referred to as `global') as the input to the AE network, and the reconstruction loss is calculated based on all events in both the original and reconstructed sequences. In local reconstruction, illustrated in Fig.~\ref{fig:local_reconstruction}, only the context of the input sequence (referred to as `local') serves as the input to the AE network, and the reconstruction loss is similarly computed using only this focus. In contrast, G2L reconstruction, shown in Fig.~\ref{fig:G2L_reconstruction}, takes the complete input sequence as its input (similar to global approaches) but computes the reconstruction loss using only the focus of the original and reconstructed sequences (akin to local approaches).

  The G2L reconstruction paradigm leverages all events (global) within a sequence to reconstruct the focus events (local), which forces the trained model focus on the reconstruction of the focus events rather than the reconstruction of all the events in sequences. Therefore, this design enables the model to enhance its learning of the normal pattern reflected by focus events during the training phase. During the inference stage, when applied to normal sequences, the insights drawn from normal focus events excel at self-reconstruction, leading to a small reconstruction error. Conversely, in the context of anomalous sequences, the presence of the abnormal focus events amplifies the reconstruction error due to the poor self-reconstruction ability of abnormal focus events. The heightened disparity in reconstruction error renders anomalous sequences more discernible under the G2L reconstruction paradigm.
  
\subsection{Three Self-supervised Tasks}
\label{subsection:self_supervised_tasks}
  LogSD employs a dual-network architecture incorporating three self-supervised tasks. In this framework, one network utilizes an auto-encoder structure, while the other follows an encoder-only design. Both networks implement their encoders and decoders using CNNs as fundamental building blocks.

  LogSD utilizes three carefully designed self-supervised tasks, i.e., reconstruction task, one-class task, and prediction task, to effectively guide model optimization during training. These tasks corresponds to three specific losses: the G2L reconstruction loss $L_{r}$, the one-class classification loss $L_{o}$, and the prediction loss $L_{p}$. 
  
  \textbf{Task I: G2L reconstruction Loss}
  Following the proposed G2L reconstruction paradigm described in Section~\ref{subsection:G2L_reconstruction_paradigm}, the reconstruction task of LogSD takes the focus $X_f$ and $\hat{X}_f$ as the inputs to generate the G2L reconstruction error $L_{r}$. As $X_f$ and $\hat{X}_f$ are derived through a context masking, elucidated in Section~\ref{subsection:frequency_based_masking_scheme}, this implies that the reconstruction task is focused on those infrequent events within the sequences. Here, the Mean Square Error(MSE) is utilized to derive the G2L reconstruction error. The loss function of reconstruction task is defined as 
  
  \begin{equation}
  \begin{aligned}
  \label{eq:rectruction_loss}
    L_{r} = \frac{1}{\vert N \vert*\vert M \vert*\vert d \vert}\sum_{i=1}^N\sum^M_{j=1}{\Vert x_{ij} - \hat{x}_{ij}\Vert^2_2} 
  \end{aligned}
  \end{equation}
  
  where $x_{ij}, \hat{x}_{ij} \in \mathbb{R}^{d}$ denotes the $d$-dimensional semantic features of the $j$-th event and its reconstructed event in $i$-th sequence of a batch, and $M$ is the total number of the focus events in the given sequence, while $N$ is the total number of log sequences in a batch. Here, $\Vert \cdot \Vert^2_2$ refers to as the squared $L_2$ norm.

\textbf{Task II: One Class Loss}
  Inspired by LogBERT~\cite{guo2021logbert} and CAT~\cite{zhang2022cat}, we also adopt an one-class classification task to govern the distribution of normal sequences representations in the latent space generated by the auto-decoder network. In order to facilitate the convergence of representations for normal sequences in the training set, we continually update the central representation, denoted as $c$. This is accomplished through a mean operation of the generated representations in each epoch during the training phase. Consequently, the optimization process entails minimizing Equation~\ref{eq:oneclass_loss} over the course of training. This leads to a desired outcome during inference, where the representation of a given normal log sequence should closely align with the center, while anomalous log sequences are expected to exhibit a substantial distance from this central reference point. Additionally, leveraging the one-class loss yields a supplementary advantage: it may aid in the convergence of the global-to-local reconstruction loss by contracting the sphere that encapsulates the latent representations of sequences. The mathematical definition of the one-class loss is as follows:

  \begin{equation}
  \begin{aligned}
  \label{eq:oneclass_loss}
    L_{o} = \frac{1}{N}\sum_{i=1}^N{\Vert z_{i} - c\Vert^2_2}
  \end{aligned}
  \end{equation}
  
  where $z_{i} \in \mathbb{R}^{l}$ denotes the $l$-dimensional representation of the i-th event sequence generated from the encoder of the AE subnet, $c$ represents the center representation of normal log sequences, calculated as $c = \frac{1}{ N}\sum^N_{i=1}{z_{i}}$, and $N$ is the total number of log sequences in a batch. 

\textbf{Task III: Prediction Loss}
  Taking inspiration from Knowledge Distillation, we employ the AE subnet within LogSD as the teacher network, while the EO subnet serves as the student. Notably, the student network takes the context $X_c$, generated by a focus masking, as input, while the teacher network operates on the input sequences $X$ without any masking. This design allows the teacher network to gather extra insights into normal patterns in comparison to the student network. Subsequently, we utilize the representations $Z_e$ extracted from the student network's latent space to predict the corresponding representations $Z_a$ generated by the teacher network. The loss for the prediction task is calculated using the Mean Square Error of predictive errors, referred to as the prediction loss. The formal definition of the prediction loss is as follows:

  \begin{equation}
  \begin{aligned}
  \label{eq:prediction_loss}
    L_{p} = \frac{1}{\vert N \vert*\vert l \vert}\sum^N_{i=1}{\Vert z_{i} - z^\prime_{i}\Vert^2_2} 
  \end{aligned}
  \end{equation}
  
  where $z_{i}, {z}_{i}^\prime \in \mathbb{R}^{l}$ denotes the $l$-dimensional representation of the $i$-th sequence generated from the encoder of the AE subnet and that from the EO subnet, and $N$ is the total number of log sequences in a batch. 

  Finally, the objective function for training LogSD is defined as shown in Equation~\ref{eq:total_loss}, where a hyper-parameter $\alpha$ is used to balance the reconstruction loss with other losses. Following the practice in~\cite{akcay2019ganomaly} and the principle of minimizing hyper-parameters, we maintain a ratio of 50:1 for the reconstruction loss based on sequences in the original space to the other losses based on the representations in the latent space, i.e, $\alpha$ is set to 50.

  \begin{equation}
  \begin{aligned}
  \label{eq:total_loss}
    {L} = \alpha{L}_{r} + {L}_{p} + {L}_{o}
  \end{aligned}
  \end{equation}

\subsection{Anomaly Detection}
\label{subsection: anoamly_detection}
   The loss value obtained from Equation~\ref{eq:prediction_loss} is utilized as the anomaly score. Following a procedure similar to training, given event sequences are input into the trained model to compute anomaly score for each sequence. These anomaly scores are then ranked, with higher scores indicating a higher likelihood of the corresponding sequences as anomalies. To evaluate LogSD, We follow the studies~\cite{wang2021multi, liu2023practical} to employ a simple threshold-moving algorithm~\cite{haibo2013imbalanced} to determine the optimal threshold. The optimal threshold would be the one resulting in the best result based on the predefined evaluation metric. In this work, we choose F1 score as the measurement to locate the threshold with the optimal balance between precision and recall.

    \begin{table*}[tb]
	  \renewcommand{\arraystretch}{0.9}
	\centering
	\caption{Overview of datasets used in the experiments.}
	\label{tab: experiment_dataset}
	\resizebox{0.9\textwidth}{!}{
	    \begin{threeparttable}
		\begin{tabular}{ lcl cccccccr}
			\toprule
			\multirow{2}{*}{Datasets} &  \multirow{2}{*}{\#Events} & \multirow{2}{*}{Window} & \multicolumn{4}{c}{Training Set (80\%)} & \multicolumn{4}{c}{Testing Set (20\%)}\\
			\cmidrule{4-11} 
			& & & \#Sequences  & \#Events & \#Norm. & \%Norm. & \#Sequences  & \#Events & \#Anom.  & \%Anom. \\
			\toprule
			HDFS& 48 & session &460,048 & 48 & 446,663 &97.07\% &115,012 &42 &3,453 &3.00\% \\
			\midrule 
			&  & 100-logs & 5,019 & 440 & 4,111 & 81.91\% & 1,254 & 226 & 129 & 10.29\% \\
			\cmidrule{3-11}
			BGL& 450& 60-logs & 8,365 & 440 & 7,110 &85.00\% & 2,091 & 226 & 150 &7.17\%\\
			\cmidrule{3-11}
			& & 20-logs & 25,095 & 440 & 22,401 & 89.26\% & 6,273 & 226 & 198 & 3.16\%\\
			\midrule
			& & 100-logs & 33,325 &1,209 & 19,744 & 59.25\% & 8,331 & 847 & 409 & 4.91\% \\
			\cmidrule{3-11}
			Spirit& 1,228& 60-logs & 55,541 & 1,209 & 34,387 &61.91\% & 13,885 & 847 & 434 & 3.13\% \\
			\cmidrule{3-11}
			& & 20-logs & 166,624 & 1,209 & 110,524 & 66.33\% & 41,656 & 847 & 465 & 1.12\%\\
			\bottomrule
		\end{tabular}
  	  \begin{tablenotes} 
		\item[1] \#Events: number of unique events; 
                  \#Sequences: number of sequences; \#Norm.: number of normal sequences; \%Norm.: percentage of normal sequences;\#Anom.: number of anomalous sequences; \%Anom.: percentage of anomalies.
            \item[2] For BGL and Spirit, consecutive duplicate events are removed from parsed logs before being grouped, which ensures that sequences with only a single unique event are eliminated, making model evaluation more realistic and reliable.
            \item[3] The data splitting strategy for BGL and Spirit follows a chronological order, with the initial 80\% of log messages used for training and the remaining 20\% for testing. This approach prevents data leakage during model training.
            \end{tablenotes}
	      \end{threeparttable} 
	}
    \end{table*}

\section{Experimental Setting}
\label{section: evaluation}

\subsection{Datasets}
  In our experiments, three public log datasets HDFS, BGL, and Spirit are used to evaluate the proposed approach and the relevant baselines. These datasets are extensively utilized in log analysis research~\cite{le2021log, zhang2020anomaly, du2017deeplog, lou2010mining, xu2009largescale, landauer2022deep} due to their nature of real-world datasets with comprehensive labeling. All these log datasets are sourced from publicly available websites~\cite{loghub_website} and ~\cite{usenix}. The details of the resulting log datasets in our experiments are summarized in Table~\ref{tab: experiment_dataset}. 

  \textbf{HDFS} dataset is generated by running Hadoop-based map-reduce jobs on more than 200 Amazon’s EC2 nodes. Given that the annotations available for this dataset are associated with the BlockId, the resultant log event sequences are organized into sessions using the BlockId. Regarding data segmentation, we adopt the methodology from~\cite{le2021log, le2022log, xie2022loggd}, allocating 80\% of the log sequences randomly for training and reserving the remaining 20\% for testing purposes.

  \textbf{BGL} dataset is an open dataset comprising logs obtained from a BlueGene/L supercomputer system at Lawrence Livermore National Labs (LLNL) in Livermore, California. This supercomputer is equipped with 131,072 processors and 32,768 GB of memory. Annotations for the BGL dataset are applied to individual log messages, resulting in the grouping of log event sequences based on entries. We specifically opt for fixed entry window grouping for the following reasons. Firstly, it enhances the generalizationability of LogSD across datasets, as not all datasets have a universally applicable identifier for session grouping. Additionally, fixed entry window grouping offers storage efficiency advantages over the entry sliding window approach. The dataset is generated under three fixed entry window size settings, namely 100 logs, 60 logs, and 20 logs. Fixed Time window grouping setting and fixed entry window grouping under larger window size settings were not chosen because they may lead to potential delays in fault discovery within real-world scenarios. For data splitting, we adhere to a chronological order strategy for the BGL dataset, wherein the initial 80\% of log messages are used for training, and the subsequent 20\% are allocated for testing. This approach aligns with real-world scenarios, ensuring that no data leakage occurs during the model training process.

  \textbf{Spirit} dataset originates from a high-performance cluster located at Sandia National Labs (SNL). This dataset was collected from systems equipped with 1,028 processors and 1,024 GB of memory. In this study, we utilize 1-gigabyte continuous log lines from the Spirit dataset for computational efficiency. The grouping strategy and data splitting approach for the Spirit dataset are identical to those used for the BGL dataset.

  It's essential to highlight that before aggregating log events into sequences, we apply a specific pre-processing step to the BGL and Spirit datasets. However, this pre-processing is not performed on the HDFS dataset since its labels are annotated at the session level, and any event removal to a sequence might affect its label. The specific pre-processing involves removing consecutive duplicate events from the parsed logs before grouping them into sequences. For instance, consider a sequence of parsed log events represented as $S = [...E_1, E_2, , E_2, , E_2, , E_1...]$. After this pre-processing, the sequence is transformed to $S = [...E_1, E_2, E_1...]$, effectively eliminating consecutive duplicates of $E_2$ that appeared in the original sequence. Subsequently, the resulting events are grouped based on entries. This pre-processing step serves a dual purpose: firstly, it reduces the overall dataset size, thus minimizing the time required for model training. Secondly, it effectively addresses potential issues with inflated detection performance, especially when a sequence consists solely of a single unique event. By removing such duplicates, the evaluation of the model's performance becomes more realistic and reliable.

\subsection{Baseline Methods}
\label{section:baseline_methods}
  To evaluate the effectiveness of the proposed method, we compare LogSD with eight state-of-the-art deep semi-supervised methods, a trivial detector, and one state-of-the-art supervised method on the aforementioned public log datasets. Specifically, AE~\cite{farzad2020unsupervised} and OC4Seq~\cite{wang2021multi} are reconstruction based and one-class classification based methods, respectively. PLElog~\cite{yang2021semi} is an approach using Probabilistic Label Estimation technique, while DeepLog/Log2Vec~\cite{meng2020semantic}, LogAnomaly~\cite{meng2019loganomaly}, Logsy~\cite{nedelkoski2020self}, LogBERT~\cite{guo2021logbert}, and CAT~\cite{zhang2022cat} belong to self-supervised methods. The random detector, as a trivial detector, randomly assigns labels to each sequence, acting as a basic baseline reference. CNN~\cite{lu2018detecting} is chosen as a representative state-of-the-art (SOTA) supervised baseline method for comparison, providing a reference to assess performance disparities between supervised and semi-supervised approaches. This selection is particularly pertinent because CNN employs the same network building blocks as LogSD, enabling a direct comparison of methodologies built upon identical fundamental components. In terms of the neural networks, these methods cover the utilization of GRU, LSTM, BiLSTM, Transformer, and CNN models, respectively, which also provide us with another perspective to examine the efficacy of different neural network building blocks.  
  
  We did not compare our method with those traditional machine learning-based methods, such as PCA~\cite{xu2009detecting}, iForest~\cite{liu2008isolation}, LogCluster~\cite{lin2016log}, OC-SVM\cite{scholkopf2001estimating}, Invariants Mining~\cite{lou2010mining}, among others. This decision was made because some studies~\cite{andonov2022logs2graphs, guo2021logbert, le2022log, zhang2022cat} had demonstrated deep learning-based methods more effective than traditional machine learning approaches in anomaly detection effectiveness. Additionally, we refrained from comparing LogSD with two GNN-based semi-supervised methods, namely GLAD-PAW~\cite{wan2021glad} and Logs2Graphs~\cite{andonov2022logs2graphs}, due to the unavailability of their implementations. We leave the comparisons with them as future work.

\subsection{Evaluation Metrics}
  To evaluate the effectiveness of the approaches, we utilize three groups of evaluation metrics. 
\begin{itemize}
  \item The first group of metrics is Matthews Correlation Coefficient (MCC)~\cite{baldi2000assessing}, which is considered as a reliable metric of the quality of prediction models~\cite{yao2021impact, lavazza2022comparing}. This metric takes into account true and false positives and negatives and is generally regarded as a balanced measure. The MCC metric is defined as: 
  $ MCC = \frac{TP \times TN - FP \times FN}{\sqrt{(TP + FP)(TP + FN)(TN + FP)(TN + FN)}} $

  \item To align with the evaluation method of prior studies~\cite{farzad2020unsupervised, wang2021multi, yang2021semi, meng2019loganomaly, meng2020semantic, nedelkoski2020self, guo2021logbert, le2022log, zhang2022cat}, therefore, precision/recall/F1 score are used as the second group of metrics in our study. Specifically, the metrics are calculated as follows:

  \textbf{Precision:} the percentage of correctly detected abnormal log sequences amongst all detected abnormal log sequences by the model.  $Precision = \frac{TP}{TP+FP}$

  \textbf{Recall:} the percentage of log sequences that are correctly identified as anomalies over 
  all real anomalies. $Recall = \frac{TP}{TP+FN}$ 

  \textbf{F1 score:} the harmonic mean of Precision and Recall. $F1 = \frac{2*Precision*Recall} {Precision+Recall}$

  \item The third group is ranking metrics, including ROC score (Area Under Receiver Operating Characteristics Curve) and PRC score (Area Under the Precision-Recall Curve). Both metrics are widely used to quantify the detection accuracy of anomaly detection~\cite{huang2005using, zhang2022cat, andonov2022logs2graphs}. Thus, we also leverage them to assess the baseline methods and LogSD. 
  
\end{itemize}

  For all the aforementioned metrics, a value closer to 1 indicates superior performance.

\begin{table*}[htbp]
  \centering
  \caption{Performance Comparison of LogSD and Baseline methods.}
  \label{tab:baseline_comparison} 
  \tiny
  \resizebox{0.9\textwidth}{!}{
      \begin{threeparttable}
      \begin{tabular}{c c c ccc ccc}
          \toprule
          Method & Metrics & HDFS & \multicolumn{3}{c}{BGL} & \multicolumn{3}{c}{Spirit} \\
          \cmidrule(lr){3-3} \cmidrule(lr){4-6} \cmidrule(lr){7-9}
          & & \textbf{session}  
            & \textbf{100-logs} & \textbf{60-logs} & \textbf{20-logs} 
            & \textbf{100-logs} & \textbf{60-logs} & \textbf{20-logs} \\
          \midrule
          \multirow{5}{*}{\textbf{AE}~\cite{farzad2020unsupervised}}  
                    & MCC & 0.7570 & 0.5764 & 0.5536 & 0.3903 & 0.1968 & 0.1651 & 0.0713 \\
                    & F1 & 0.7432 & 0.6165 & 0.5475 & 0.3911 & 0.2101 & 0.1631 & 0.0719 \\
                    & Precision  & 0.8445 & 0.7450 & 0.5090 & 0.3129 & 0.1214 & 0.0970 & 0.0549 \\
                    & Recall  & 0.6735 & 0.5297 & 0.5933 & 0.5236 & 0.7889 & 0.5922 & 0.1835 \\
                    & PRC & 0.6175 & 0.5974 & 0.4697 & 0.2489 & 0.1112 & 0.0800 & 0.0247 \\
                    & ROC & 0.7191 & 0.8189 & 0.8623 & 0.8898 & 0.7924 & 0.7653 & 0.5963 \\
          \midrule
          \multirow{5}{*}{\textbf{OC4Seq}~\cite{wang2021multi}} 
                    & MCC & 0.9287 & 0.7528 & 0.6939 & 0.4809 & 0.4217 & 0.3268 & 0.1232 \\
                    & F1 & 0.9315 & 0.7894 & 0.6859 & 0.4764 & 0.4074 & 0.3069 & 0.1257 \\
                    & Precision  & 0.9362 & 0.7944 & 0.6254 & 0.3986 & 0.2754 & 0.1890 & 0.0812 \\
                    & Recall  & 0.9268 & 0.7959 & 0.7711 & 0.7138 & 0.8354 & 0.8487 & 0.2875 \\
                    & PRC & \underline{0.9582} & 0.7117 & 0.5522 & 0.4390 & 0.2416 & 0.1634 & 0.0423 \\
                    & ROC & \underline{0.9988} & \underline{0.9634} & \underline{0.9604} & 0.9617 & 0.9168 & 0.9142 & 0.7539 \\
          \midrule
          \multirow{5}{*}{\textbf{PLELog}~\cite{yang2021semi}} 
                    & MCC & \underline{0.9447} & \underline{0.8122} & \underline{0.7557} & \underline{0.7817} & \underline{0.5283} & \underline{0.4349} & \underline{0.4943} \\
                    & F1 & \underline{0.9451} & \underline{0.8249} & \underline{0.7684} & \underline{0.7853} & \underline{0.4903} & 0.3664 & \underline{0.4192} \\
                    & Precision  & 0.9617 & 0.7441 & 0.6468 & 0.7637 & 0.3322 & 0.2302 & \underline{0.2722} \\
                    & Recall  & \underline{0.9299} & \underline{0.9259} & \textbf{0.9477} & 0.8157 & \underline{0.9544} & \underline{0.9531} & \textbf{0.9384} \\
                    & PRC & 0.8949 & 0.6969 & \underline{0.6167} & 0.6275 & \underline{0.3193} & \underline{0.2207} & \underline{0.2559} \\
                    & ROC & 0.9647 & 0.9452 &          0.9534 & 0.9036 & \underline{0.9254} & \underline{0.9196} & \underline{0.9542} \\
          \midrule
          \multirow{5}{*}{\textbf{DeepLog/}\textbf{Log2Vec}~\cite{meng2020semantic}}  
                    & MCC & 0.9393 & 0.6059 & 0.5606 & 0.4573 & 0.1126 & 0.0962 & 0.1321 \\
                    & F1 & 0.9295 & 0.6358 & 0.5889 & 0.4543 & 0.1643 & 0.1261 & 0.1398 \\
                    & Precision  & \underline{0.9716} & 0.6767 & 0.5169 & 0.3845 & 0.1073 & 0.1243 & 0.1400 \\
                    & Recall  & 0.8911 & 0.6006 & 0.6856 & 0.5567 & 0.3537 & 0.1289 & 0.1395  \\
                    & PRC & 0.8691 & 0.6569 & 0.6121 & 0.4776 & 0.0696 & 0.0432 & 0.0292 \\
                    & ROC & 0.9452 & 0.7858 & 0.8190 & 0.7639 & 0.6004 & 0.5497 & 0.5650 \\
          \cmidrule{3-9}
          \multirow{5}{*}{\textbf{LogAnomaly}~\cite{meng2019loganomaly}} 
                    & MCC & 0.7524 & 0.6320 & 0.4283 & 0.3313 & 0.0737 & 0.0853 & 0.0796 \\          
                    & F1 & 0.7333 & 0.6404 & 0.4515 & 0.3116 & 0.0797 & 0.0860 & 0.0908 \\
                    & Precision  & \textbf{0.9985} & 0.6952 & 0.3314 & 0.2102 & 0.1245 & 0.1214 & 0.0758 \\ 
                    & Recall     & 0.5856          & 0.5938 & 0.7136 & 0.6531 & 0.0587 & 0.0668 & 0.1133 \\
                    & PRC  & 0.6390          & 0.5208 & 0.3817 & 0.2438 & 0.1003 & 0.0373 & 0.0185 \\
                    & ROC  & 0.7927          & 0.8909 & 0.8686 & 0.8689 & 0.7274 & 0.5255 & 0.5488 \\
          \cmidrule{3-9}
          \multirow{5}{*}{\textbf{Logsy}~\cite{nedelkoski2020self}} 
                    & MCC & 0.9156 & 0.6827 & 0.6033 & 0.4796 & 0.1823 & 0.1453 & 0.1310 \\          
                    & F1 & 0.9225 & 0.7155 & 0.6282 & 0.4735 & 0.1873 & 0.1536 & 0.1337 \\ 
                    & Precision & 0.9405 & 0.6630 & 0.5762 & 0.3879 & 0.1094 & 0.1196 & 0.1242 \\          
                    & Recall & 0.9054 & 0.7855 & 0.6978 & 0.6094 & 0.6504 & 0.2220 & 0.1455 \\
                    & PRC & 0.8549 & 0.5415 & 0.4222 & 0.2498 & 0.0883 & 0.0506 & 0.0276 \\
                    & ROC & 0.9518 & 0.8697 & 0.8288 & 0.7889 & 0.6886 & 0.5841 & 0.5669 \\
          \cmidrule{3-9}
          \multirow{5}{*}{\textbf{LogBERT}~\cite{guo2021logbert}} 
                    & MCC & 0.7473 & 0.7419 & 0.6782 & 0.6533 & 0.4117 & 0.3905 & 0.3771 \\          
                    & F1 & 0.7255 & 0.7646 & 0.6873 & 0.6652 & 0.4266 & \underline{0.3939} & 0.3535 \\
                    & Precision  & 0.9435 & \underline{0.8408} & 0.5983 & 0.5830 & 0.3236 & \underline{0.2950} & 0.2544 \\
                    & Recall  & 0.5894 & 0.7028 & \underline{0.8578} & 0.7828 & 0.6259 & 0.5929 & 0.5792 \\
                    & PRC & 0.4794 & \underline{0.8314} & 0.5601 & 0.5035 & 0.1975 & 0.1752 & 0.1340 \\
                    & ROC & 0.9370 & 0.9624 & 0.9337 & 0.9472 & 0.8662 & 0.8870 & 0.9187 \\
          \cmidrule{3-9}
          \multirow{5}{*}{\textbf{CAT}~\cite{zhang2022cat}}            
                    & MCC & 0.6235 & 0.6861 & 0.4987 & 0.2745 & 0.3276 & 0.2513 & 0.0511 \\ 
                    & F1 & 0.6458 & 0.7164 & 0.4835 & 0.2942 & 0.3248 & 0.2535 & 0.0537 \\
                    & Precision  & 0.7909 & 0.7396 & \underline{0.7677} & 0.2100 & 0.5059 & 0.2438 & 0.0492 \\
                    & Recall  & 0.5457 & 0.6810 & 0.4167 & 0.5069 & 0.3238 & 0.3220 & 0.0789 \\
                    & PRC & 0.5124 & 0.5679 & 0.4132 & 0.1441 & 0.2320 & 0.1315 & 0.0131 \\
                    & ROC & 0.8164 & 0.8257 & 0.7058 & 0.7319 & 0.7482 & 0.6845 & 0.3779 \\
          \midrule
          \multirow{5}{*}{\textbf{LogSD}}    
                    & MCC & \textbf{0.9559} & \textbf{0.9483} & \textbf{0.8712} & \textbf{0.9047} & \textbf{0.8954} & \textbf{0.8714} & \textbf{0.8676} \\          
                    & F1 & \textbf{0.9583} & \textbf{0.9627} & \textbf{0.8664} & \textbf{0.8906} & \textbf{0.8957} & \textbf{0.8861} & \textbf{0.8707} \\ 
                    & Precision  & 0.9587 & \textbf{0.9600} & \textbf{0.8753} & \textbf{0.9118} & \textbf{0.8386} & \textbf{0.8287} & \textbf{0.9209} \\
                    & Recall  & \textbf{0.9580} & \textbf{0.9664} & \underline{0.8578} & \textbf{0.8704} & \textbf{0.9650} & \textbf{0.9616} & \underline{0.8258} \\
                    & PRC & \textbf{0.9840} & \textbf{0.9716} & \textbf{0.8736} & \textbf{0.9416} & \textbf{0.7346} & \textbf{0.7609} & \textbf{0.8013} \\
                    & ROC & \textbf{0.9993} & \textbf{0.9977} & \textbf{0.9911} & \textbf{0.9967} & \textbf{0.9927} & \textbf{0.9961} & \textbf{0.9987} \\
          \bottomrule
      \end{tabular}%
      \begin{tablenotes} 
          \tiny\itshape  
          \item[1] All the models were trained for up to 100 epochs and an early stopping strategy was used for 20 consecutive iterations without performance improvement. 
          \item[2] The figures in the table that were underlined represent the second-best metrics, while those in bold font indicated the best metrics.
          \item[3] Each method was executed three times for each dataset configuration, and the resulting values were averaged to report the final results.
      \end{tablenotes}
      \end{threeparttable} 
  }
\end{table*}%

\subsection{Implementation and Environment}
  In the implementation of LogSD, we chose the state-of-the-art tool, Drain \cite{he2017drain} for log parsing because of its proven effectiveness, robustness, and efficiency~\cite{zhu2019tools, dai2023pilar, fu2023empirical}. We implemented LogSD in PyTorch 1.11.0. The encoder and decoder in LogSD are single-layer 2D-convolution and 2D-transposed convolution modules, respectively. Following the practice in ~\cite{chen2021experience}, we extracted a 32-dimensional vector as the semantic embedding for each log event. The hidden dimension of all convolution networks in LogSD was set to 128, with three kernel sizes specified as [3, 4, 5]. LogSD was trained using the AdamW optimizer with polynomial learning rate decay at $1\times10^{-4}$. The learning rate started at $1\times10^{-2}$ and gradually decreased to $1\times10^{-4}$. We set the mini-batch size to 64. The model underwent training for up to 100 epochs, with an early stopping strategy applied when performance failed to improve for 20 consecutive iterations. The hyper-parameter $\kappa$ for the frequency-based masking ratio was dynamically selected from the set $K = \{0.05, 0.1, 0.15, 0.2, 0.3\}$. Additionally, the hyper-parameter for the reconstruction loss weight, $\alpha$, was set to 50. 

  In our comparative analysis, the implementations of baseline approaches were obtained from public repositories~\cite{DeepLoglizercode, OC4Seqcode, PLELogcode, LogAnomalycode, LBcode, CATcode}.
  For consistency, we used the parameters provided by their respective implementations unless specified otherwise. In our experiments, each method was executed three times for each dataset configuration, and the resulting values were averaged to report the final results. All experiments were conducted on a Linux server Ubuntu 20.04 equipped with an AMD Ryzen 3.5GHz CPU, 96 GB of memory, and an RTX2080Ti with 11 GB of GPU memory.  

\afterpage{%

\begin{table*}[htbp]
  \centering
  \caption{Performance Comparison of LogSD and Extra Baseline Methods.}
  \label{tab:extra_baseline_comparison} 
  \tiny
  \resizebox{0.9\textwidth}{!}{
      \begin{threeparttable}
      \begin{tabular}{c c c ccc ccc}
          \toprule
          Method & Metrics & HDFS & \multicolumn{3}{c}{BGL} & \multicolumn{3}{c}{Spirit} \\
          \cmidrule(lr){3-3} \cmidrule(lr){4-6} \cmidrule(lr){7-9}
          & & \textbf{session}  
            & \textbf{100-logs} & \textbf{60-logs} & \textbf{20-logs} 
            & \textbf{100-logs} & \textbf{60-logs} & \textbf{20-logs} \\
        \midrule
        \multirow{5}{*}{\textbf{Random Detector}}    
            & MCC & 0.0018& 0.0170& 0.0210& 0.0232& 0.0055& 0.0022& 0.0088\\ 
            & F1 & 0.0994& 0.1690& 0.1330& 0.0683& 0.0916& 0.0594& 0.0244\\
            & Precision & 0.0552& 0.1155& 0.0787& 0.0400& 0.0510& 0.0319& 0.0130\\ 
            & Recall & 0.4998& 0.3643& 0.4333& 0.2643& 0.4613& 0.4424& 0.2968\\
            & PRC & 0.0552& 0.1122& 0.0747& 0.0335& 0.0498& 0.0320& 0.0115\\
            & ROC & 0.4996& 0.4909& 0.5049& 0.5052& 0.4998& 0.4958& 0.5157\\
          \midrule
          \multirow{5}{*}
          {\textbf{CNN~\cite{lu2018detecting}}} 
            & MCC & 0.9803& 0.9746& 0.9645& 0.9744& 0.9825& 0.9758& 0.9719\\
            & F1 & 0.9807& 0.9768& 0.9671& 0.9760& 0.9828& 0.9760& 0.9718\\
            & Precision & 0.9664& 0.9725& 0.9652& 0.9697& 0.9965& 0.9975& 0.9961\\          
            & Recall & 0.9956& 0.9812& 0.9695& 0.9825& 0.9695& 0.9554& 0.9487\\    
            & PRC & 0.9562& 0.9558& 0.9380& 0.9538& 0.9669& 0.9537& 0.9453\\
            & ROC & 0.9972& 0.9893& 0.9833& 0.9902& 0.9847& 0.9777& 0.9743\\
          \midrule
          \multirow{5}{*}{\textbf{LogSD}}    
                    & MCC & 0.9559 & 0.9483 & 0.8712 & 0.9047 & 0.8954 & 0.8714 & 0.8676 \\          
                    & F1 & 0.9583 & 0.9627 & 0.8664 & 0.8906 & 0.8957 & 0.8861 & 0.8707 \\ 
                    & Precision  & 0.9587 & 0.9600 & 0.8753 & 0.9118 & 0.8386 & 0.8287 & 0.9209 \\
                    & Recall  & 0.9580 & 0.9664 & 0.8578 & 0.8704 & 0.9650 & 0.9616 & 0.8258 \\
                    & PRC & 0.9840 & 0.9716 & 0.8736 & 0.9416 & 0.7346 & 0.7609 & 0.8013 \\
                    & ROC & 0.9993 & 0.9977 & 0.9911 & 0.9967 & 0.9927 & 0.9961 & 0.9987 \\
          \bottomrule
      \end{tabular}%
      \end{threeparttable} 
  }
\end{table*}%
}

\section{Results and Analysis}
\label{section: result_and_analysis}

\subsection{Comparison with State-of-the-art Methods}

  In this experiment, our objective is to assess the effectiveness of LogSD through a comparative study with baseline methods on the three aforementioned public log datasets. To ensure a fair comparison, all methods that utilize semantic embeddings as inputs, including AE, DeepLog/Log2Vec, Logsy, and LogSD (excluding PLELog and CAT), employ the identical semantic extraction scheme as detailed in~\cite{chen2021experience}. PLELog and CAT maintain their original semantic scheme. LogAnomaly, OC4Seq, and LogBERT utilize log indices as input, without any modifications to their original implementations. 
  
  The experimental results, as presented in Table~\ref{tab:baseline_comparison},  consistently highlight the superior performance of LogSD across three datasets with varying window settings, in terms of the MCC, F1 score and ranking metrics. Notably, LogSD achieves superior results on the HDFS dataset, outperforming all the baseline methods that also demonstrate commendable performance. On the BGL and Spirit datasets, LogSD significantly outperforms the baseline methods. 
  In terms of F1-score, it achieves an improvement of 16.7\%, 12.7\%, and 13.4\% under the 100-logs, 60-logs, and 20-logs settings on the BGL dataset over the second-best baseline method, and remarkable improvements of 82.6\%, 141.8\%, and 107.7\% under the Spirit dataset's corresponding settings. These results provide empirical evidence for the overall effectiveness of our proposed LogSD. In contrast, PLELog 
  secures the second-best performance across the three datasets in most settings. This observation suggests that combining probabilistic label estimation with supervised learning can be a promising solution for anomaly detection in semi-supervised scenarios. Moreover, LogBERT exhibits better performance on the BGL and Spirit datasets compared to other baselines, specifically outperforming the other self-supervised approaches~\cite{meng2019loganomaly, meng2020semantic, nedelkoski2020self, zhang2022cat} working under the predefined positional masking scheme and local reconstruction paradigm described in Section~\ref{section:related_work}. Despite this, it is still far inferior to our LogSD that combines the G2L reconstruction paradigm with frequency-based masking scheme. In addition, the reconstruction-based method, AE, exhibits lower performance compared to the one-class classification-based method, OC4Seq. This observation implies that global reconstruction may limit the model's capacity to effectively differentiate anomalies from normal samples.

  When considering the selection of neural networks, the results in Table~\ref{tab:baseline_comparison} demonstrate that our CNN-based LogSD outperforms all baseline methods using other neural networks,including AE (BiLSTM), OC4Seq (LSTM), PLELog (LSTM), DeepLog/Log2Vec (LSTM), LogAnomaly (LSTM), Logsy (Transformer), LogBERT (Transformer), and CAT (Transformer). This solidifies the advantage of CNN networks, consistent with findings in previous studies~\cite{lu2018detecting, chen2021experience, le2022log}. Another noteworthy observation pertains to the methods categorized under the self-supervised type. The three methods based on Transformer networks---Logsy, CAT, and LogBERT---consistently demonstrate superior performance compared to LSTM-based methods like DeepLog/Log2Vec and LogAnomaly in most cases. This observation strengthens the perception of Transformer networks holding an advantage over LSTM networks. However, upon examining methods across different technique categories, drawing deterministic conclusions from their performance comparison becomes intricate. The LSTM-based methods, OC4Seq and PLELog, occasionally outperform the Transformer-based methods, Logsy, CAT, and LogBERT. This suggests that while the choice of neural network types does influence detection performance, its impact may be less pronounced compared to selecting appropriate anomaly detection techniques among reconstruction, one-class classification, probabilistic label estimation, and self-supervised learning.

  Additionally, we also compare LogSD with a random detector and a SOTA supervised method, CNN~\cite{lu2018detecting}. The experimental results in Table~\ref{tab:extra_baseline_comparison}, reveal that LogSD significantly outperforms the random detector in all experimental settings. While the CNN method outperforms LogSD, it's important to note that CNN is a supervised method, whereas LogSD operates as a semi-supervised method. This indicates that the additional prior knowledge employed in the supervised method enhances the detection performance, particularly when there is an ample amount of labeled data available.

\begin{table*}[htbp]
  \centering
  \caption{Ablation Experiments for Network, Masking Schemes, and Reconstruction Paradigms.}
  \label{tab:ablation_experiments} 
  \tiny
  \resizebox{0.95\textwidth}{!}{
      \begin{threeparttable}
      \begin{tabular}{c cccc cccc cccc}
          \toprule
          Dataset & \multicolumn{4}{c}{HDFS} & \multicolumn{4}{c}{BGL} & \multicolumn{4}{c}{Spirit} \\
          \cmidrule(lr){2-5} \cmidrule(lr){6-9} \cmidrule(lr){10-13}
          & \textbf{MCC} & \textbf{F1} & \textbf{PRC} & \textbf{ROC}  
          & \textbf{MCC} & \textbf{F1} & \textbf{PRC} & \textbf{ROC}  
          & \textbf{MCC} & \textbf{F1} & \textbf{PRC} & \textbf{ROC} \\
          \midrule
          LogSD\textsubscript{sng}  & 0.7880 & 0.7858 & 0.7345 & 0.8793 & 0.9126 & 0.9152 & 0.9474 & 0.9935 & 0.7801 & 0.7949 & 0.7074 & 0.9781 \\
          \cmidrule{2-13}
          LogSD\textsubscript{srl} & 0.3872 & 0.3895 & 0.3071 & 0.5999 & 0.8357 & 0.8226 & 0.8336 & 0.9209 & 0.3335 & 0.3340 & 0.2590 & 0.6771 \\
          \cmidrule{2-13}
          LogSD\textsubscript{sfl} & 0.7122 & 0.7143 & 0.6583 & 0.8468 & 0.9095 & 0.8939 & 0.9377 & 0.9921 & 0.8769 & 0.8846 & \textbf{0.7731} & 0.9634 \\
          \cmidrule{2-13}
          LogSD\textsubscript{srf} & 0.9101 & 0.9067 & 0.9019 & 0.9977 & 0.9264 & 0.9339 & 0.9616 & 0.9965 & 0.8032 & 0.8012 & 0.7055 & 0.9824 \\
          \cmidrule{2-13}
          LogSD\textsubscript{sff} & 0.9213 & 0.9153 & 0.9403 & 0.9954 & \textbf{0.9596} & 0.9534 & 0.9644 & \textbf{0.9977} & 0.8930 & 0.8921 & 0.7509 & 0.9837 \\
          \midrule
          LogSD\textsubscript{dng} & 0.9471& 0.9462 & 0.9821 & \textbf{0.9995} & 0.9384 & 0.9366 & 0.9489 & 0.9905 & 0.8913 & 0.8936 & 0.7343 & 0.9925 \\
          \cmidrule{2-13}
          LogSD\textsubscript{drl} & 0.4223 & 0.4120 & 0.3450 & 0.5979 & 0.8382 & 0.8371 & 0.8631 & 0.9515 & 0.3577 & 0.3420 & 0.2669 & 0.7270 \\
          \cmidrule{2-13}
          LogSD\textsubscript{dfl} & 0.7688 & 0.7597 & 0.7473& 0.9458 & 0.9418 & 0.9281 & 0.9489 & 0.9905 & 0.8812 & 0.8886 & 0.7368 & 0.9833 \\
          \cmidrule{2-13}
          LogSD\textsubscript{drf} & 0.9319 & 0.9338 & 0.9812& \textbf{0.9995} & 0.9329 & 0.9498 & 0.9697 & 0.9974 & 0.8807 & 0.8911 & 0.6930 & 0.9914\\
          \cmidrule{2-13}
          \textbf{LogSD\textsubscript{dff}} & \textbf{0.9559} & \textbf{0.9583} & \textbf{0.9840}& 0.9993 & 0.9483 & \textbf{0.9627} & \textbf{0.9716} & \textbf{0.9977} & \textbf{0.8954} &\textbf{0.8957} & 0.7346 & \textbf{0.9927}\\    
          \bottomrule
      \end{tabular}%
      \begin{tablenotes} 
          \tiny\itshape  
          \item[1] Notes: the first character in the subscript signifies the network type, where "s" denotes a single network, and "d" indicates a dual-network.
          The second character represents the masking type, where "n" implies no masking, "r" stands for randomly masking, and "f" corresponds to frequency-based masking.
          The third character indicates the reconstruction type, with "g" standing for global reconstruction, "l" indicating local reconstruction, and "f" representing global-to-local reconstruction. Notably, LogSD\textsubscript{dff} is identical to LogSD, bearing the name LogSD\textsubscript{dff} for the purpose of convenient correspondence with its counterpart within a single network group.
      \end{tablenotes}
      \end{threeparttable}
    }
 \end{table*}%

\subsection{Ablation Studies}
  In this section, our objective is to comprehensively examine the effectiveness of each major component within LogSD on the final results. To achieve this, we conduct ablation studies that categorize the LogSD variants into three distinct groups:
  \begin{itemize}
    \item \textbf{Network Comparison:} We compare the variants with and without a dual-network architecture.
    \item \textbf{Masking Scheme Comparison:} We evaluate the performance of variants with different masking schemes: no masking, randomly masking, and frequency-based masking.
    \item \textbf{Reconstruction Paradigm Comparison:} We assess and contrast the variants with distinct reconstruction paradigms, including global, local, and G2L reconstruction paradigms.
  \end{itemize}

  Due to space constraints, we only present experimental results for the HDFS session setting and both BGL and Spirit datasets under the 100-logs window setting. The results are shown in Table~\ref{tab:ablation_experiments}.

\subsubsection{Single-network Vs. Dual-network}
  From Table~\ref{tab:ablation_experiments}, it is evident that LogSD variants employing dual-network (subscripts starting with "d") consistently outperform their counterparts with a single network (subscripts starting with "s"). These results empirically confirm our hypothesis that integrating a self-supervised prediction task between dual networks in LogSD enhances the model's capability to learn patterns from normal sequences, aiding in better distinguishing anomalies from normal sequences.

\subsubsection{No masking Vs. Randomly Masking Vs. Frequency-Based Masking}
  Table~\ref{tab:ablation_experiments} provides a clear comparison between LogSD variants denoted by "$n$" (for no masking), "$r$" (for randomly masking) and "$f$" (for frequency-based masking) across different reconstruction paradigms, using either single or dual networks. The results consistently demonstrate the superiority of frequency-based masking in terms of F1-score. Moreover, it's evident that LogSD variants with random masking exhibit significantly higher performance fluctuations compared to their frequency-based masking counterparts. This variance suggests a potential drawback associated with the randomly masking scheme. Overall, the experimental results empirically support our claim that frequency-based masking facilitates the trained model in effectively learning normal patterns through the reconstruction of low-frequency events. Consequently, when faced with new sequences, our approach excels in distinguishing these anomalies from normal sequences, particularly those with rare normal events.

\subsubsection{Global Vs. Local Vs. G2L reconstruction} 
  Furthermore, we investigate the performance difference of three reconstruction paradigm on log anomaly detection. From Table\ref{tab:ablation_experiments}, we can see that those LogSD variants subscript ending with "$f$" outperform their counterparts with subscript ending with "$l$" or "$g$" in most cases. Another interesting thing we can see is that those methods under local reconstruction paradigm are not necessarily superior to those under global reconstruction paradigm. This may imply that feeding masked inputs into reconstruction networks in the local reconstruction paradigm carries a potential risk of information loss, which leads to sub-optimal detection performance for the corresponding variant methods. Conversely, the global-to-local paradigm might be able to avoid this shortcoming in the local reconstruction paradigm.

\begin{figure}
\centering
\subfloat[Loss Weight Sensitivity.]{\label{fig:loss_weight} \includegraphics[width=0.48\textwidth]{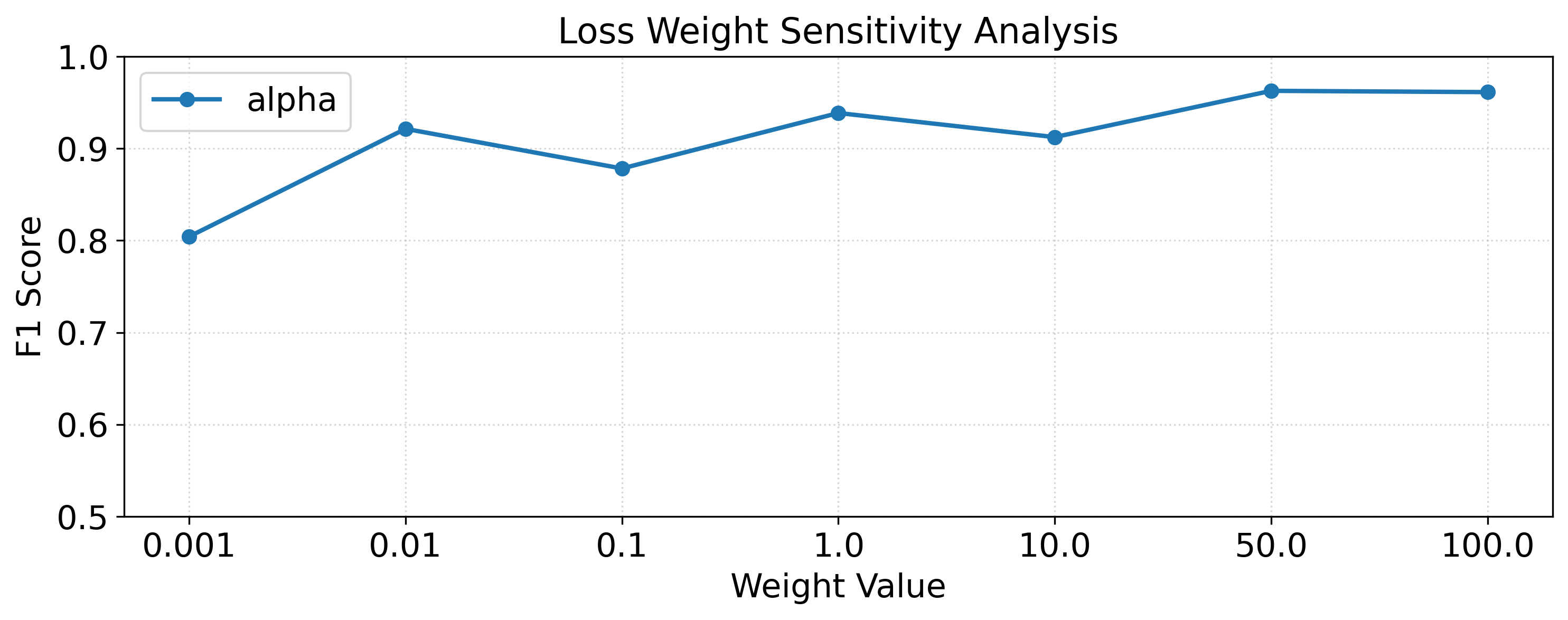}}
\subfloat[Masking Ratio Sensitivity.]{\label{fig:masking-ratio} \includegraphics[width=0.48\textwidth]{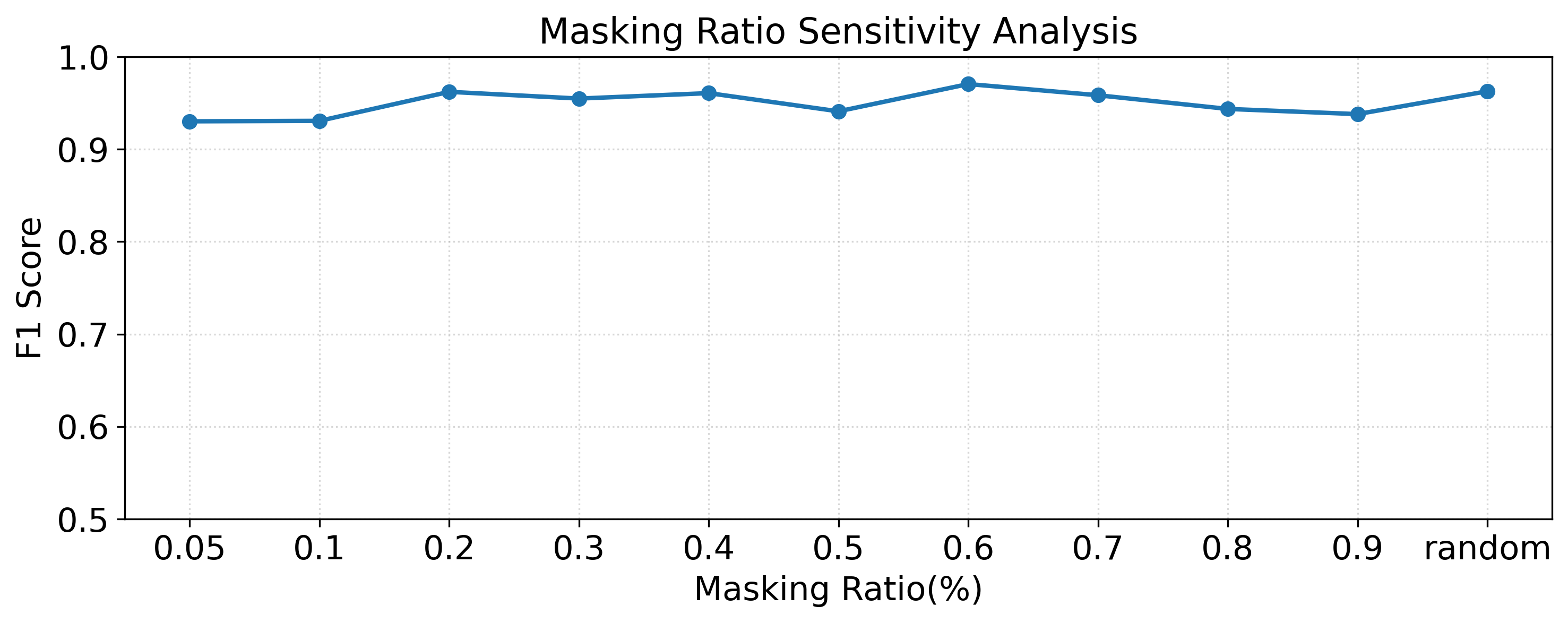}}%
\caption{Sensitivity analysis on BGL under 100 logs window setting.}
\label{fig: sensitivity_analysis}
\end{figure}

\subsubsection{The impact of consecutive duplicate events preprocessing} 
  Two types of sequences in the original datasets may influence the performance of LogSD. The first is the sequences containing only a single unique event. The presence of such sequences can undermine LogSD's frequency-based masking mechanism, diminishing the model's effectiveness in learning infrequent norm event patterns. The second is the sequences with consecutive duplicate events but more than one event. These sequences introduce higher fluctuations in the distribution of events, posing challenges in applying a specific masking ratio or a small random masking ratio set that ensures competitive model performance across diverse datasets. The experimental results for the BGL and Spirit datasets, categorized by original sequences, sequences without single events, and preprocessed sequences, are provided in Table~\ref{tab:preprocessing_experiments}. The results clearly indicate that the removal of consecutive duplicate events leads to an enhancement in the performance of LogSD.

\afterpage{%

\begin{table*}[htbp]
  \centering
  \caption{The Impact of Data Preprocessing on the Performance of LogSD.}
  \label{tab:preprocessing_experiments} 
  \tiny
  \resizebox{0.8\textwidth}{!}{
    \begin{tabular}{c c ccc ccc}
      \toprule
      Dataset & Metric &\multicolumn{3}{c}{BGL} & \multicolumn{3}{c}{Spirit} \\
      \cmidrule(lr){3-5} \cmidrule(lr){6-8} 
      & &\textbf{100-logs} & \textbf{60-logs} & \textbf{20-logs} & 
         \textbf{100-logs} & \textbf{60-logs} & \textbf{20-logs} \\
      \midrule
                     & MCC & 0.7166 & 0.5817 & 0.6611 & 0.8934 & 0.8638 & 0.7381 \\
                     & F1  & 0.7385 & 0.5861 & 0.6587 & 0.8912 & 0.8588 & 0.7086 \\
        Original     & Precision & 0.7371 & 0.5957 & 0.8062 & 0.8128 & 0.7605 & 0.5503 \\
        Sequences    & Recall    & 0.7549 & 0.7002 & 0.5866 & 0.9882 & 0.9862 & 0.9947 \\
                     & PRC & 0.6751 & 0.4490 & 0.5348 & 0.7542 & 0.6610 & 0.7512 \\
                     & ROC & 0.9180 & 0.8816 & 0.8808 & 0.9968 & 0.9968 & 0.9992 \\
       \midrule
                     & MCC & 0.8238 & 0.7933 & 0.6741 & 0.8934 & 0.8639 & 0.7949 \\
        No           & F1  & 0.8413 & 0.8087 & 0.6811 & 0.8919 & 0.8675 & 0.7789 \\
        Single-event & Precision & 0.8885 & 0.8296 & 0.7504 & 0.8128 & 0.7700 & 0.6444 \\
        Sequences    & Recall    & 0.7998 & 0.7971 & 0.6658 & 0.9882 & 0.9838 & 0.9842 \\
                     & PRC & 0.8222 & 0.6892 & 0.5597 & 0.7580 & 0.7709 & 0.7331 \\
                     & ROC & 0.9297 & 0.9136 & 0.9109 & 0.9967 & 0.9979 & 0.9990 \\
       \midrule
                     & MCC       & 0.9483 & 0.8712 & 0.9047 & 0.8954 & 0.8714 & 0.8676 \\
                     & F1        & 0.9627 & 0.8664 & 0.8906 & 0.8957 & 0.8861 & 0.8707 \\
       Preprocessed  & Precision & 0.9600 & 0.8753 & 0.9118 & 0.8386 & 0.8287 & 0.9209 \\
       Sequences     & Recall    & 0.9664 & 0.8578 & 0.8704 & 0.9650 & 0.9616 & 0.8258 \\
                     & PRC       & 0.9716 & 0.8736 & 0.9416 & 0.7346 & 0.7609 & 0.8013 \\
                     & ROC       & 0.9977 & 0.9911 & 0.9967 & 0.9927 & 0.9961 & 0.9987 \\
       
      \bottomrule
    \end{tabular}%
  }

\end{table*}%
}

\subsection{Sensitivity Studies}
\subsubsection{Loss Weights Sensitivity} In the previous experiments, all assessments were conducted with the hyper-parameter value set at $\alpha=50$. To delve deeper into the impact of these weights, a sensitivity analysis was conducted. The hyper-parameter was varied within the range of $10^{-3}$ to $10^{2}$. As illustrated in Figure ~\ref{fig:loss_weight}, the influence of hyper-parameter choices for the loss weight on the model's performance varies from 0.8043 to 0.9627. Specifically, increasing the weight $\alpha$ for the reconstruction loss generally corresponds to higher accuracy, peaking at 50. Therefore, we empirically opt to maintain the current hyper-parameter setting.

\subsubsection{Masking Ratio Sensitivity} Regarding the frequency-based masking ratio hyper-parameter, we investigated its influence on detection performance. This involved setting the masking ratio to a constant value ranging from 0.05 to 0.9, as well as utilizing a random value chosen from our empirically determined set $K =\{0.05, 0.1, 0.15, 0.2, 0.3\}$. From Figure~\ref{fig:masking-ratio}, we observe a slight impact of random masking ratios on LogSD's performance, attributed to event distribution characteristics within a specified window size. In the comparison between fixed masking ratios and randomly chosen ratios from $K$, the latter revealed commendable results, albeit not the optimal. Opting for a unified set of randomly chosen ratios mitigates challenges in hyper-parameter selection for diverse datasets, leading us to adopt this approach in this study.

\begin{figure}
\centering
\subfloat[Training Time.]{\label{train_time} \includegraphics[width=0.48\textwidth]{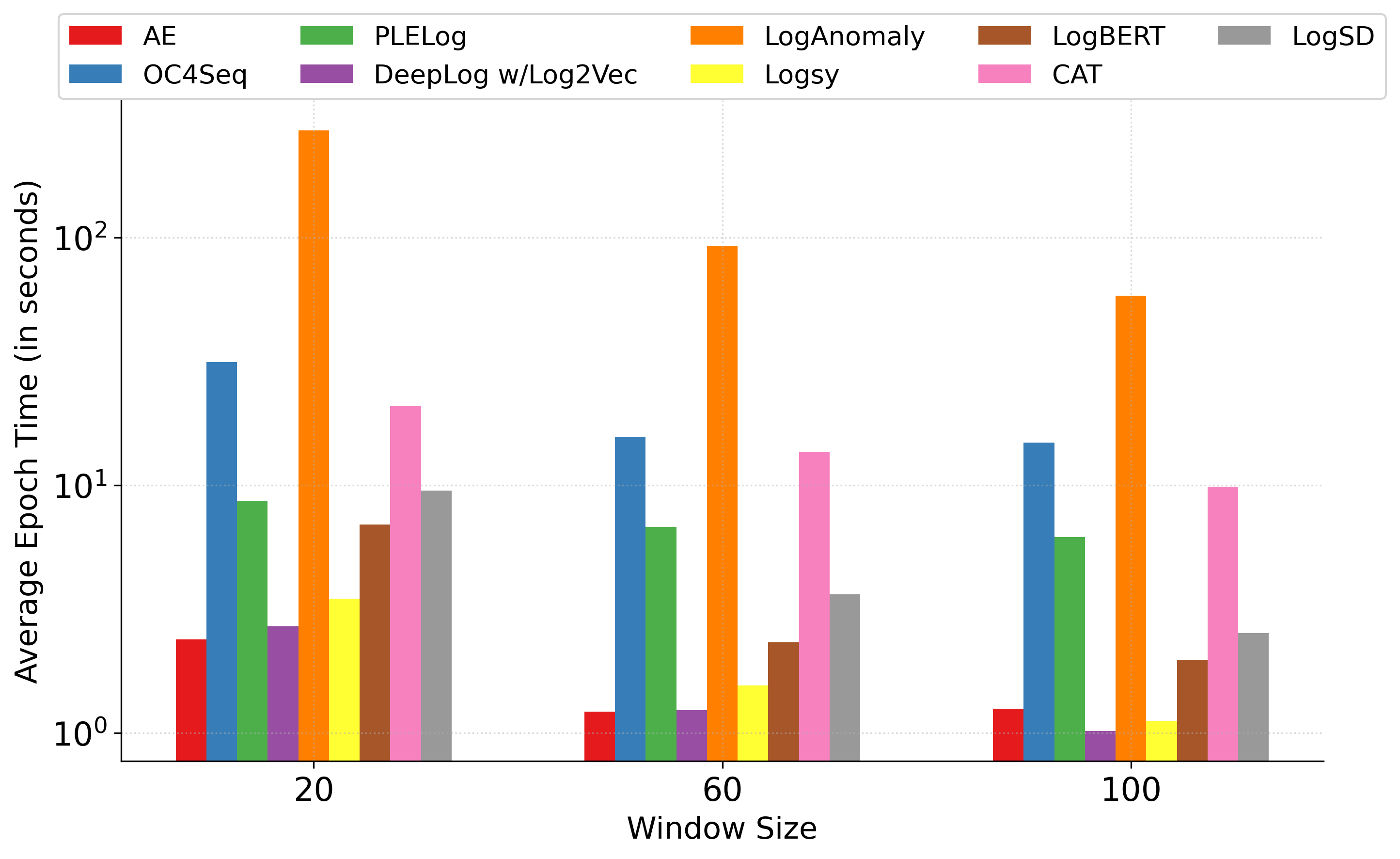}}%
\hfill
\subfloat[Testing Time.]{\label{test_time} \includegraphics[width=0.48\textwidth]{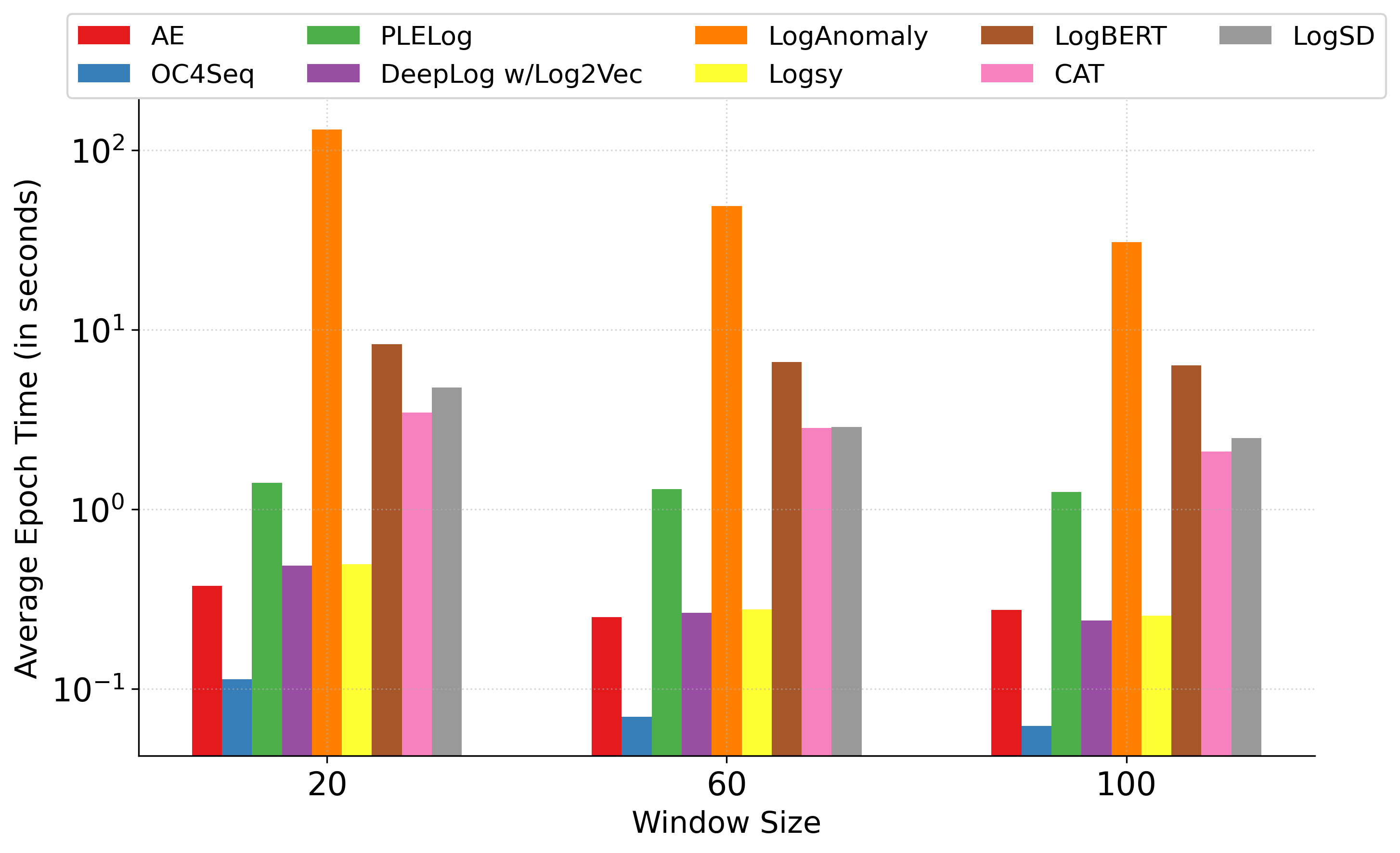}}%
\caption{Scalability analysis on BGL under 100 logs window setting.}
\label{fig:scalability_analysis}
\end{figure}

\subsection{Scalability Studies}
To evaluate the scalability of LogSD, we conducted runtime comparisons with the baseline methods, showcasing the efficiency of our CNN-based LogSD. Figure~\ref{fig:scalability_analysis} presents the average training and testing times per epoch for these approaches across varying window size settings. Due to a consistent trend in time efficiency, we just present the experimental results on the BGL dataset with a window size of 100 logs for conciseness. LogSD demonstrates comparable training efficiency in comparison to the baseline methods. In terms of testing time, LogSD is shorter than LogAnomaly and LogBERT, similar to CAT, yet slightly longer than other baseline methods. 

\section{Discussion}
\subsection{The Usability of the Proposed Method}
In the log analysis domain, beyond the effectiveness of detection, the usability of a detection method also plays a pivotal role in real-world scenarios. The usability of a detection method typically encompass: integration capability and cost with existing software systems generating logs, adaptability to system evolution, and root cause localization capability. (1) LogSD showcases robust integration capabilities as it doesn't depend on any prerequisite prior knowledge about the system generating the logs. LogSD can perform well in both session-based and time-window-based datasets, allowing it to adapt effectively to different blackbox applications. (2) Regarding the data collection, preprocessing and model training cost, LogSD is similar to existing methods except an additional preprocessing step to remove consecutive events, whose time cost is negligible. Therefore, it is not costly to integrate LogSD with an existing system. (3) In terms of adaptability, regular model retraining is a usual essential strategy when significant changes occur in the business workflow of a system. The efficient model training/retraining time, depicted in Figure~\ref{fig:scalability_analysis}, contributes to LogSD's adaptability in dynamic environments. (4) Root cause analysis is crucial for operators to reduce the impact of a system failure. Our proposed method, incorporating a masking mechanism, can be enhanced for fine-grained root cause analysis by integrating a progressive masking refinement mechanism. We will leave it to the future work.

\subsection{Threats To Validity}
\label{subsection:thread_to_validity}
  In this study, we identify the following threats to validity:

  \textbf{External Threats:} This threat pertains to factors affecting the generalization of results. One potential external threat to validity of this study is the subject selection bias introduced by the choice of datasets. We utilized three log datasets for our experiments, all of them are drawn from real-world industrial systems and are prevalent in existing literature \cite{xie2021logdp,le2021log, le2022log, li2023graph}. 
  Our future work will encompass a more diverse range of datasets and systems to further evaluate the generalizability of our approach.

  \textbf{Internal Threats:} This refers to potential threats that might affect the results and have not been adequately considered. In our study, the internal threat to validity primarily lies in the implementations of LogSD and the compared approaches, as well as the design of experiments. To mitigate the threat stemming from the implementation, we conducted rigorous inspection and testing of the program. We assumed that all bugs revealed during testing were fixed in our approach. LogSD was implemented based on popular libraries, and we made our source code available for replication purposes. Concerning the compared methods, we utilized their open-source implementations. To address potential threats related to the experimental design, we took several measures. First, we ran all experiments three times and reported the averaged results. Second, we ensured a consistent setting for all methods, such as using 100 epochs for training and a patience of 20 for early stopping. Additionally, for the baselines, we selected the learning rate and relevant configurations as per the best-reported value in their respective papers. We believe this careful experimental design facilitates fair comparisons among all evaluated methods.

\section{Related Work}
\label{section:related_work}

\subsection{Existing Masking Schemes}
\label{subsection:existing_masking_scheme}
  The utilization of masking as a data augmentation technique has gained widespread recognition in many self-supervised learning 
  based log anomaly detection methods~\cite{du2017deeplog, meng2019loganomaly, nedelkoski2020self, guo2021logbert, zhang2022cat}. In these methods, a masking scheme serves as one of pre-processing steps to enable the trained model to utilize the unmasked event to reconstruct/predict the masked events for log anomaly detection. The masking schemes in existing log anomaly detection methods can be roughly categorized into predefined positional masking scheme and randomly masking scheme. Predefined positional masking scheme involves the selection of specific positional indices where events occur within a sequence for input sequence event masking. Typically, these indices are chosen in advance and remain unchanged throughout the model training and inference processes. 
  The methods using predefined positional masking scheme include ~\cite{du2017deeplog, meng2019loganomaly,nedelkoski2020self, zhang2022cat}. In contrast, the positional indices in randomly masking scheme are selected on the fly in a random manner, which makes sequence event masking more flexible and diverse. ~\cite{guo2021logbert} is the representative of the method using this masking scheme. However, both of the aforementioned masking schemes fail to take the occurrence frequency of events in sequences into consideration, making it challenging for the model to effectively distinguish between anomalous sequences and normal sequences that comprise infrequent events.
  
\subsection{Existing Reconstruction Paradigms}
\label{subsection:existing_reconstruction_paradigm}
  The reconstruction paradigm in existing methods can be classified into: global reconstruction paradigm and local reconstruction paradigm. Some existing reconstruction based methods like AE~\cite{farzad2020unsupervised} often employ the following reconstruction paradigm: sequences are fed into a reconstruction network featuring an auto-encoder architecture. Subsequently, reconstruction errors are computed by comparing the reconstructed sequences with the original ones. Given that the reconstruction covers all events within a sequence, we refer to it as the "global reconstruction" paradigm. As overemphasizing the reconstruction of each event within sequences, the global reconstruction paradigm may hinder model convergence, especially when dealing with sequences exhibiting complex distributions~\cite{xu2023fascinating}. Consequently, it may lead to sub-optimal outcomes for those methods using the global reconstruction paradigm. 
  
  The local reconstruction paradigm involves inputting sequences with intentionally masked specific events into a reconstruction/prediction network. Subsequently, the masked events are reconstructed using the unmasked events within the sequences. We name this paradigm as local reconstruction paradigm because they leverage one portion of events in sequences to reconstruct the other portion. This paradigm is widely employed in many models like BERT~\cite{devlin2018bert} and Masked Autoencoder~\cite{he2022masked}. In the log analysis domain, this paradigm is also widely used in many studies, such as DeepLog/LogVec, LogAnomaly, Logsy, CAT, and LogBERT. However, the local reconstruct paradigm does not fully exploit the potential information contained within the masked events themselves to generate informative indicators for anomaly detection. This limitation may result in sub-optimal performance in the context of log sequence anomaly detection. 

\subsection{Existing Deep Semi-supervised Approaches to Log-based Anomaly Detection}

 From the perspective of techniques, existing deep semi-supervised log-based anomaly detection approaches can be broadly classified as follows: 

\begin{itemize}  

  \item{One class classification-based approach}, which commonly utilizes machine learning or deep learning techniques to encapsulate normal data or their latent representations within a compact space. Then, any given instance positioned beyond the boundaries of the learned compact space is classified as anomalous. OC4Seq~\cite{wang2021multi} is the representative of this approach. Nonetheless, the effectiveness of one-class models can be compromised when confronted with datasets exhibiting intricate distributions within normal samples~\cite{huang2022self, xu2023fascinating}. 
  
  \item{Probabilistic label estimation-based approach}, which employs the idea of Positive and Unlabeled Learning (PU Learning)~\cite{elkan2008learning} to initially derive pseudo labels for given unlabeled instances based on known normal log sequences in the training set. Subsequently, a supervised deep learning model is trained on these data with pseudo labels to classify these instances as anomalous or normal. A notable exemplar within this approach is PLELog~\cite{yang2021semi}. However, the effectiveness of this approach is intrinsically linked to the accuracy of probabilistic label estimation. The estimation outcome is, in turn, influenced by the inherent characteristics of the data and the techniques employed for probabilistic label estimation. 
  
  \item{Reconstruction-based approach} works on the assumption that a normal data instance can be well reconstructed from its low-dimensional latent representation, while anomalies exhibit higher reconstruction errors compared to their normal counterparts. This idea is utilized in the method of AE~\cite{farzad2020unsupervised}. However, this approach still struggles to achieve satisfactory performance due to its overemphasis on reconstructing the entire input~\cite{xu2023fascinating}. 
  
  \item{Self-supervised learning-based approach}, which leverages designed self-supervised tasks to steer model optimization~\cite{xie2022self}, consequently yielding specialized representations for log-based anomaly detection. DeepLog~\cite{du2017deeplog}, LogBERT~\cite{guo2021logbert} and CAT~\cite{zhang2022cat} serve as notable exemplars of this approach. However, as mentioned in Section~\ref{subsection:existing_masking_scheme} and ~\ref{subsection:existing_reconstruction_paradigm}, the existing masking schemes and the reconstruction paradigm may potentially impede their models' capacity to learn patterns from normal instances effectively. This could lead to a compromise in the overall detection accuracy.
  
\end{itemize}

\section{Conclusion}
\label{section: conclusion}
  In this paper, we have proposed LogSD, a novel semi-supervised log-based anomaly detection approach. LogSD encompasses a frequency-based masking scheme, a G2L reconstruction paradigm, and a dual-network framework with three self-supervised learning tasks. The frequency-based masking scheme empowers the model to emphasize the learning of normal patterns associated with infrequent events. The global-to-local reconstruction paradigm capitalizes on all the events in sequences for the reconstruction of those infrequent events. The dual-network framework with three self-supervised tasks not only bolsters the model by leveraging information from both the original space and the latent space but amalgamates the strengths of reconstruction, one-class-classification, and knowledge distillation techniques. Through the collaboration of these components, LogSD efficiently captures less biased and highly discriminative normal patterns from infrequent log messages. The acquired less biased normal patterns enable LogSD to better discriminate between anomalous sequences and those containing sporadic events. Our extensive experiments on three widely-used public datasets illustrate the effectiveness of LogSD. 

  Our source code and experimental data are available at~\textbf{\url{https://github.com/ilwoof/logsd/}}.
 

\begin{acks}
  This research was supported by an Australian Government Research Training Program (RTP) Scholarship, and by the Australian Research Council’s Discovery Projects funding scheme (project DP200102940). The work was also supported with super-computing resources provided by the Phoenix High Powered Computing (HPC) service at the University of Adelaide.
\end{acks}

\bibliographystyle{ACM-Reference-Format}
\bibliography{paper-reference}


\begin{thebibliography}{69}


\ifx \showCODEN    \undefined \def \showCODEN     #1{\unskip}     \fi
\ifx \showDOI      \undefined \def \showDOI       #1{#1}\fi
\ifx \showISBNx    \undefined \def \showISBNx     #1{\unskip}     \fi
\ifx \showISBNxiii \undefined \def \showISBNxiii  #1{\unskip}     \fi
\ifx \showISSN     \undefined \def \showISSN      #1{\unskip}     \fi
\ifx \showLCCN     \undefined \def \showLCCN      #1{\unskip}     \fi
\ifx \shownote     \undefined \def \shownote      #1{#1}          \fi
\ifx \showarticletitle \undefined \def \showarticletitle #1{#1}   \fi
\ifx \showURL      \undefined \def \showURL       {\relax}        \fi
\providecommand\bibfield[2]{#2}
\providecommand\bibinfo[2]{#2}
\providecommand\natexlab[1]{#1}
\providecommand\showeprint[2][]{arXiv:#2}

\bibitem[Log(2019)]%
        {LogAnomalycode}
 \bibinfo{year}{2019}\natexlab{}.
\newblock \bibinfo{title}{{LogAnomaly} Code Repository}.
\newblock
  \bibinfo{howpublished}{\url{https://github.com/donglee-afar/logdeep}}.
\newblock


\bibitem[Dee(2021)]%
        {DeepLoglizercode}
 \bibinfo{year}{2021}\natexlab{}.
\newblock \bibinfo{title}{{DeepLoglizer} Code Repository}.
\newblock
  \bibinfo{howpublished}{\url{https://github.com/logpai/deep-loglizer}}.
\newblock


\bibitem[LBc(2021)]%
        {LBcode}
 \bibinfo{year}{2021}\natexlab{}.
\newblock \bibinfo{title}{{LogBert} Code Repository}.
\newblock \bibinfo{howpublished}{\url{https://github.com/HelenGuohx/logbert}}.
\newblock


\bibitem[OC4(2021)]%
        {OC4Seqcode}
 \bibinfo{year}{2021}\natexlab{}.
\newblock \bibinfo{title}{{OC4Seq} Code Repository}.
\newblock
  \bibinfo{howpublished}{\url{https://github.com/wzwtrevor/Multi-Scale-One-Class-Recurrent-Neural-Networks}}.
\newblock


\bibitem[PLE(2021)]%
        {PLELogcode}
 \bibinfo{year}{2021}\natexlab{}.
\newblock \bibinfo{title}{{PLELog} Code Repository}.
\newblock \bibinfo{howpublished}{\url{https://github.com/LeonYang95/PLELog}}.
\newblock


\bibitem[CAT(2022)]%
        {CATcode}
 \bibinfo{year}{2022}\natexlab{}.
\newblock \bibinfo{title}{{CAT} Code Repository}.
\newblock \bibinfo{howpublished}{\url{https://github.com/mmichaelzhang/CAT}}.
\newblock


\bibitem[Akcay et~al\mbox{.}(2019)]%
        {akcay2019ganomaly}
\bibfield{author}{\bibinfo{person}{Samet Akcay}, \bibinfo{person}{Amir
  Atapour-Abarghouei}, {and} \bibinfo{person}{Toby~P Breckon}.}
  \bibinfo{year}{2019}\natexlab{}.
\newblock \showarticletitle{Ganomaly: Semi-supervised anomaly detection via
  adversarial training}. In \bibinfo{booktitle}{\emph{Computer Vision--ACCV
  2018: 14th Asian Conference on Computer Vision, Perth, Australia, December
  2--6, 2018, Revised Selected Papers, Part III 14}}. Springer,
  \bibinfo{pages}{622--637}.
\newblock


\bibitem[Andonov et~al\mbox{.}(2022)]%
        {andonov2022logs2graphs}
\bibfield{author}{\bibinfo{person}{Stefan Andonov}, \bibinfo{person}{Viktor
  Jovev}, \bibinfo{person}{Aleksandar Kitanovski}, \bibinfo{person}{Aleksandar
  Krsteski}, {and} \bibinfo{person}{Gjorgji Madjarov}.}
  \bibinfo{year}{2022}\natexlab{}.
\newblock \showarticletitle{logs2graphs: Data-driven graph representation and
  visualization of log data}.
\newblock


\bibitem[Baldi et~al\mbox{.}(2000)]%
        {baldi2000assessing}
\bibfield{author}{\bibinfo{person}{Pierre Baldi}, \bibinfo{person}{S{\o}ren
  Brunak}, \bibinfo{person}{Yves Chauvin}, \bibinfo{person}{Claus~AF Andersen},
  {and} \bibinfo{person}{Henrik Nielsen}.} \bibinfo{year}{2000}\natexlab{}.
\newblock \showarticletitle{Assessing the accuracy of prediction algorithms for
  classification: an overview}.
\newblock \bibinfo{journal}{\emph{Bioinformatics}} \bibinfo{volume}{16},
  \bibinfo{number}{5} (\bibinfo{year}{2000}), \bibinfo{pages}{412--424}.
\newblock


\bibitem[Bodik et~al\mbox{.}(2010)]%
        {bodik2010fingerprinting}
\bibfield{author}{\bibinfo{person}{Peter Bodik}, \bibinfo{person}{Moises
  Goldszmidt}, \bibinfo{person}{Armando Fox}, \bibinfo{person}{Dawn~B Woodard},
  {and} \bibinfo{person}{Hans Andersen}.} \bibinfo{year}{2010}\natexlab{}.
\newblock \showarticletitle{Fingerprinting the datacenter: automated
  classification of performance crises}. In
  \bibinfo{booktitle}{\emph{Proceedings of the 5th European conference on
  Computer systems}}. \bibinfo{pages}{111--124}.
\newblock


\bibitem[Chen et~al\mbox{.}(2004)]%
        {chen2004failure}
\bibfield{author}{\bibinfo{person}{Mike Chen}, \bibinfo{person}{Alice~X Zheng},
  \bibinfo{person}{Jim Lloyd}, \bibinfo{person}{Michael~I Jordan}, {and}
  \bibinfo{person}{Eric Brewer}.} \bibinfo{year}{2004}\natexlab{}.
\newblock \showarticletitle{Failure diagnosis using decision trees}. In
  \bibinfo{booktitle}{\emph{International Conference on Autonomic Computing,
  2004. Proceedings.}} IEEE, \bibinfo{pages}{36--43}.
\newblock


\bibitem[Chen et~al\mbox{.}(2021)]%
        {chen2021experience}
\bibfield{author}{\bibinfo{person}{Zhuangbin Chen}, \bibinfo{person}{Jinyang
  Liu}, \bibinfo{person}{Wenwei Gu}, \bibinfo{person}{Yuxin Su}, {and}
  \bibinfo{person}{Michael~R Lyu}.} \bibinfo{year}{2021}\natexlab{}.
\newblock \showarticletitle{Experience Report: Deep Learning-based System Log
  Analysis for Anomaly Detection}.
\newblock \bibinfo{journal}{\emph{arXiv preprint arXiv:2107.05908}}
  (\bibinfo{year}{2021}).
\newblock


\bibitem[Dai et~al\mbox{.}(2023)]%
        {dai2023pilar}
\bibfield{author}{\bibinfo{person}{Hetong Dai}, \bibinfo{person}{Yiming Tang},
  \bibinfo{person}{Heng Li}, {and} \bibinfo{person}{Weiyi Shang}.}
  \bibinfo{year}{2023}\natexlab{}.
\newblock \showarticletitle{PILAR: Studying and Mitigating the Influence of
  Configurations on Log Parsing}. In \bibinfo{booktitle}{\emph{2023 IEEE/ACM
  45th International Conference on Software Engineering (ICSE)}}. IEEE,
  \bibinfo{pages}{818--829}.
\newblock


\bibitem[Devlin et~al\mbox{.}(2018)]%
        {devlin2018bert}
\bibfield{author}{\bibinfo{person}{Jacob Devlin}, \bibinfo{person}{Ming-Wei
  Chang}, \bibinfo{person}{Kenton Lee}, {and} \bibinfo{person}{Kristina
  Toutanova}.} \bibinfo{year}{2018}\natexlab{}.
\newblock \showarticletitle{Bert: Pre-training of deep bidirectional
  transformers for language understanding}.
\newblock \bibinfo{journal}{\emph{arXiv preprint arXiv:1810.04805}}
  (\bibinfo{year}{2018}).
\newblock


\bibitem[Du et~al\mbox{.}(2017)]%
        {du2017deeplog}
\bibfield{author}{\bibinfo{person}{Min Du}, \bibinfo{person}{Feifei Li},
  \bibinfo{person}{Guineng Zheng}, {and} \bibinfo{person}{Vivek Srikumar}.}
  \bibinfo{year}{2017}\natexlab{}.
\newblock \showarticletitle{Deeplog: Anomaly detection and diagnosis from
  system logs through deep learning}. In \bibinfo{booktitle}{\emph{Proceedings
  of the 2017 ACM SIGSAC conference on computer and communications security}}.
  \bibinfo{pages}{1285--1298}.
\newblock


\bibitem[Elkan and Noto(2008)]%
        {elkan2008learning}
\bibfield{author}{\bibinfo{person}{Charles Elkan} {and} \bibinfo{person}{Keith
  Noto}.} \bibinfo{year}{2008}\natexlab{}.
\newblock \showarticletitle{Learning classifiers from only positive and
  unlabeled data}. In \bibinfo{booktitle}{\emph{Proceedings of the 14th ACM
  SIGKDD international conference on Knowledge discovery and data mining}}.
  \bibinfo{pages}{213--220}.
\newblock


\bibitem[Farzad and Gulliver(2020)]%
        {farzad2020unsupervised}
\bibfield{author}{\bibinfo{person}{Amir Farzad} {and} \bibinfo{person}{T~Aaron
  Gulliver}.} \bibinfo{year}{2020}\natexlab{}.
\newblock \showarticletitle{Unsupervised log message anomaly detection}.
\newblock \bibinfo{journal}{\emph{ICT Express}} \bibinfo{volume}{6},
  \bibinfo{number}{3} (\bibinfo{year}{2020}), \bibinfo{pages}{229--237}.
\newblock


\bibitem[Fu et~al\mbox{.}(2023)]%
        {fu2023empirical}
\bibfield{author}{\bibinfo{person}{Ying Fu}, \bibinfo{person}{Meng Yan},
  \bibinfo{person}{Zhou Xu}, \bibinfo{person}{Xin Xia},
  \bibinfo{person}{Xiaohong Zhang}, {and} \bibinfo{person}{Dan Yang}.}
  \bibinfo{year}{2023}\natexlab{}.
\newblock \showarticletitle{An empirical study of the impact of log parsers on
  the performance of log-based anomaly detection}.
\newblock \bibinfo{journal}{\emph{Empirical Software Engineering}}
  \bibinfo{volume}{28}, \bibinfo{number}{1} (\bibinfo{year}{2023}),
  \bibinfo{pages}{6}.
\newblock


\bibitem[Guo et~al\mbox{.}(2021)]%
        {guo2021logbert}
\bibfield{author}{\bibinfo{person}{Haixuan Guo}, \bibinfo{person}{Shuhan Yuan},
  {and} \bibinfo{person}{Xintao Wu}.} \bibinfo{year}{2021}\natexlab{}.
\newblock \showarticletitle{Logbert: Log anomaly detection via bert}. In
  \bibinfo{booktitle}{\emph{2021 international joint conference on neural
  networks (IJCNN)}}. IEEE, \bibinfo{pages}{1--8}.
\newblock


\bibitem[Haibo and Yunqian(2013)]%
        {haibo2013imbalanced}
\bibfield{author}{\bibinfo{person}{He Haibo} {and} \bibinfo{person}{Ma
  Yunqian}.} \bibinfo{year}{2013}\natexlab{}.
\newblock \showarticletitle{Imbalanced learning: foundations, algorithms, and
  applications}.
\newblock \bibinfo{journal}{\emph{Wiley-IEEE Press}} \bibinfo{volume}{1},
  \bibinfo{number}{27} (\bibinfo{year}{2013}), \bibinfo{pages}{12}.
\newblock


\bibitem[He et~al\mbox{.}(2022)]%
        {he2022masked}
\bibfield{author}{\bibinfo{person}{Kaiming He}, \bibinfo{person}{Xinlei Chen},
  \bibinfo{person}{Saining Xie}, \bibinfo{person}{Yanghao Li},
  \bibinfo{person}{Piotr Doll{\'a}r}, {and} \bibinfo{person}{Ross Girshick}.}
  \bibinfo{year}{2022}\natexlab{}.
\newblock \showarticletitle{Masked autoencoders are scalable vision learners}.
  In \bibinfo{booktitle}{\emph{Proceedings of the IEEE/CVF conference on
  computer vision and pattern recognition}}. \bibinfo{pages}{16000--16009}.
\newblock


\bibitem[He et~al\mbox{.}(2017)]%
        {he2017drain}
\bibfield{author}{\bibinfo{person}{Pinjia He}, \bibinfo{person}{Jieming Zhu},
  \bibinfo{person}{Zibin Zheng}, {and} \bibinfo{person}{Michael~R Lyu}.}
  \bibinfo{year}{2017}\natexlab{}.
\newblock \showarticletitle{Drain: An online log parsing approach with fixed
  depth tree}. In \bibinfo{booktitle}{\emph{2017 IEEE international conference
  on web services (ICWS)}}. IEEE, \bibinfo{pages}{33--40}.
\newblock


\bibitem[Hinton et~al\mbox{.}(2015)]%
        {hinton2015distilling}
\bibfield{author}{\bibinfo{person}{Geoffrey Hinton}, \bibinfo{person}{Oriol
  Vinyals}, {and} \bibinfo{person}{Jeff Dean}.}
  \bibinfo{year}{2015}\natexlab{}.
\newblock \showarticletitle{Distilling the knowledge in a neural network}.
\newblock \bibinfo{journal}{\emph{arXiv preprint arXiv:1503.02531}}
  (\bibinfo{year}{2015}).
\newblock


\bibitem[Huang et~al\mbox{.}(2022)]%
        {huang2022self}
\bibfield{author}{\bibinfo{person}{Chaoqin Huang}, \bibinfo{person}{Qinwei Xu},
  \bibinfo{person}{Yanfeng Wang}, \bibinfo{person}{Yu Wang}, {and}
  \bibinfo{person}{Ya Zhang}.} \bibinfo{year}{2022}\natexlab{}.
\newblock \showarticletitle{Self-supervised masking for unsupervised anomaly
  detection and localization}.
\newblock \bibinfo{journal}{\emph{IEEE Transactions on Multimedia}}
  (\bibinfo{year}{2022}).
\newblock


\bibitem[Huang and Ling(2005)]%
        {huang2005using}
\bibfield{author}{\bibinfo{person}{Jin Huang} {and} \bibinfo{person}{Charles~X
  Ling}.} \bibinfo{year}{2005}\natexlab{}.
\newblock \showarticletitle{Using AUC and accuracy in evaluating learning
  algorithms}.
\newblock \bibinfo{journal}{\emph{IEEE Transactions on knowledge and Data
  Engineering}} \bibinfo{volume}{17}, \bibinfo{number}{3}
  (\bibinfo{year}{2005}), \bibinfo{pages}{299--310}.
\newblock


\bibitem[Joulin et~al\mbox{.}(2016)]%
        {joulin2016fasttext}
\bibfield{author}{\bibinfo{person}{Armand Joulin}, \bibinfo{person}{Edouard
  Grave}, \bibinfo{person}{Piotr Bojanowski}, \bibinfo{person}{Matthijs Douze},
  \bibinfo{person}{H{\'e}rve J{\'e}gou}, {and} \bibinfo{person}{Tomas
  Mikolov}.} \bibinfo{year}{2016}\natexlab{}.
\newblock \showarticletitle{Fasttext. zip: Compressing text classification
  models}.
\newblock \bibinfo{journal}{\emph{arXiv preprint arXiv:1612.03651}}
  (\bibinfo{year}{2016}).
\newblock


\bibitem[Landauer et~al\mbox{.}(2022)]%
        {landauer2022deep}
\bibfield{author}{\bibinfo{person}{Max Landauer}, \bibinfo{person}{Sebastian
  Onder}, \bibinfo{person}{Florian Skopik}, {and} \bibinfo{person}{Markus
  Wurzenberger}.} \bibinfo{year}{2022}\natexlab{}.
\newblock \showarticletitle{Deep Learning for Anomaly Detection in Log Data: A
  Survey}.
\newblock \bibinfo{journal}{\emph{arXiv preprint arXiv:2207.03820}}
  (\bibinfo{year}{2022}).
\newblock


\bibitem[Lavazza and Morasca(2022)]%
        {lavazza2022comparing}
\bibfield{author}{\bibinfo{person}{Luigi Lavazza} {and} \bibinfo{person}{Sandro
  Morasca}.} \bibinfo{year}{2022}\natexlab{}.
\newblock \showarticletitle{Comparing $\phi$ and the F-measure as performance
  metrics for software-related classifications}.
\newblock \bibinfo{journal}{\emph{Empirical Software Engineering}}
  \bibinfo{volume}{27}, \bibinfo{number}{7} (\bibinfo{year}{2022}),
  \bibinfo{pages}{185}.
\newblock


\bibitem[Le and Zhang(2021)]%
        {le2021log}
\bibfield{author}{\bibinfo{person}{Van-Hoang Le} {and} \bibinfo{person}{Hongyu
  Zhang}.} \bibinfo{year}{2021}\natexlab{}.
\newblock \showarticletitle{Log-based anomaly detection without log parsing}.
  In \bibinfo{booktitle}{\emph{2021 36th IEEE/ACM International Conference on
  Automated Software Engineering (ASE)}}. IEEE, \bibinfo{pages}{492--504}.
\newblock


\bibitem[Le and Zhang(2022)]%
        {le2022log}
\bibfield{author}{\bibinfo{person}{Van~Hoang Le} {and} \bibinfo{person}{Hongyu
  Zhang}.} \bibinfo{year}{2022}\natexlab{}.
\newblock \showarticletitle{Log-based Anomaly Detection with Deep Learning: How
  Far Are We?}
\newblock \bibinfo{journal}{\emph{arXiv preprint arXiv:2202.04301}}
  (\bibinfo{year}{2022}).
\newblock


\bibitem[Li et~al\mbox{.}(2023a)]%
        {li2023they}
\bibfield{author}{\bibinfo{person}{Zhenhao Li}, \bibinfo{person}{An~Ran Chen},
  \bibinfo{person}{Xing Hu}, \bibinfo{person}{Xin Xia},
  \bibinfo{person}{Tse-Hsun Chen}, {and} \bibinfo{person}{Weiyi Shang}.}
  \bibinfo{year}{2023}\natexlab{a}.
\newblock \showarticletitle{Are They All Good? Studying Practitioners'
  Expectations on the Readability of Log Messages}.
\newblock \bibinfo{journal}{\emph{arXiv preprint arXiv:2308.08836}}
  (\bibinfo{year}{2023}).
\newblock


\bibitem[Li et~al\mbox{.}(2023b)]%
        {li2023graph}
\bibfield{author}{\bibinfo{person}{Zhong Li}, \bibinfo{person}{Jiayang Shi},
  {and} \bibinfo{person}{Matthijs van Leeuwen}.}
  \bibinfo{year}{2023}\natexlab{b}.
\newblock \showarticletitle{Graph Neural Network based Log Anomaly Detection
  and Explanation}.
\newblock \bibinfo{journal}{\emph{arXiv preprint arXiv:2307.00527}}
  (\bibinfo{year}{2023}).
\newblock


\bibitem[Liang et~al\mbox{.}(2007)]%
        {liang2007failure}
\bibfield{author}{\bibinfo{person}{Yinglung Liang}, \bibinfo{person}{Yanyong
  Zhang}, \bibinfo{person}{Hui Xiong}, {and} \bibinfo{person}{Ramendra Sahoo}.}
  \bibinfo{year}{2007}\natexlab{}.
\newblock \showarticletitle{Failure prediction in ibm bluegene/l event logs}.
  In \bibinfo{booktitle}{\emph{Seventh IEEE International Conference on Data
  Mining (ICDM 2007)}}. IEEE, \bibinfo{pages}{583--588}.
\newblock


\bibitem[Liao et~al\mbox{.}(2020)]%
        {liao2020using}
\bibfield{author}{\bibinfo{person}{Lizhi Liao}, \bibinfo{person}{Jinfu Chen},
  \bibinfo{person}{Heng Li}, \bibinfo{person}{Yi Zeng}, \bibinfo{person}{Weiyi
  Shang}, \bibinfo{person}{Jianmei Guo}, \bibinfo{person}{Catalin Sporea},
  \bibinfo{person}{Andrei Toma}, {and} \bibinfo{person}{Sarah Sajedi}.}
  \bibinfo{year}{2020}\natexlab{}.
\newblock \showarticletitle{Using black-box performance models to detect
  performance regressions under varying workloads: an empirical study}.
\newblock \bibinfo{journal}{\emph{Empirical Software Engineering}}
  \bibinfo{volume}{25} (\bibinfo{year}{2020}), \bibinfo{pages}{4130--4160}.
\newblock


\bibitem[Lin et~al\mbox{.}(2016)]%
        {lin2016log}
\bibfield{author}{\bibinfo{person}{Qingwei Lin}, \bibinfo{person}{Hongyu
  Zhang}, \bibinfo{person}{Jian-Guang Lou}, \bibinfo{person}{Yu Zhang}, {and}
  \bibinfo{person}{Xuewei Chen}.} \bibinfo{year}{2016}\natexlab{}.
\newblock \showarticletitle{Log clustering based problem identification for
  online service systems}. In \bibinfo{booktitle}{\emph{2016 IEEE/ACM 38th
  International Conference on Software Engineering Companion (ICSE-C)}}. IEEE,
  \bibinfo{pages}{102--111}.
\newblock


\bibitem[Liu et~al\mbox{.}(2008)]%
        {liu2008isolation}
\bibfield{author}{\bibinfo{person}{Fei~Tony Liu}, \bibinfo{person}{Kai~Ming
  Ting}, {and} \bibinfo{person}{Zhi-Hua Zhou}.}
  \bibinfo{year}{2008}\natexlab{}.
\newblock \showarticletitle{Isolation forest}. In
  \bibinfo{booktitle}{\emph{2008 eighth ieee international conference on data
  mining}}. IEEE, \bibinfo{pages}{413--422}.
\newblock


\bibitem[Liu et~al\mbox{.}(2023)]%
        {liu2023practical}
\bibfield{author}{\bibinfo{person}{Jinyang Liu}, \bibinfo{person}{Tianyi Yang},
  \bibinfo{person}{Zhuangbin Chen}, \bibinfo{person}{Yuxin Su},
  \bibinfo{person}{Cong Feng}, \bibinfo{person}{Zengyin Yang}, {and}
  \bibinfo{person}{Michael~R Lyu}.} \bibinfo{year}{2023}\natexlab{}.
\newblock \showarticletitle{Practical Anomaly Detection over Multivariate
  Monitoring Metrics for Online Services}.
\newblock \bibinfo{journal}{\emph{arXiv preprint arXiv:2308.09937}}
  (\bibinfo{year}{2023}).
\newblock


\bibitem[Lou et~al\mbox{.}(2010)]%
        {lou2010mining}
\bibfield{author}{\bibinfo{person}{Jian-Guang Lou}, \bibinfo{person}{Qiang Fu},
  \bibinfo{person}{Shengqi Yang}, \bibinfo{person}{Ye Xu}, {and}
  \bibinfo{person}{Jiang Li}.} \bibinfo{year}{2010}\natexlab{}.
\newblock \showarticletitle{Mining Invariants from Console Logs for System
  Problem Detection.}. In \bibinfo{booktitle}{\emph{USENIX Annual Technical
  Conference}}. \bibinfo{pages}{1--14}.
\newblock


\bibitem[Lu et~al\mbox{.}(2018)]%
        {lu2018detecting}
\bibfield{author}{\bibinfo{person}{Siyang Lu}, \bibinfo{person}{Xiang Wei},
  \bibinfo{person}{Yandong Li}, {and} \bibinfo{person}{Liqiang Wang}.}
  \bibinfo{year}{2018}\natexlab{}.
\newblock \showarticletitle{Detecting anomaly in big data system logs using
  convolutional neural network}. In \bibinfo{booktitle}{\emph{2018 IEEE 16th
  Intl Conf on Dependable, Autonomic and Secure Computing}}. IEEE,
  \bibinfo{pages}{151--158}.
\newblock


\bibitem[Ma et~al\mbox{.}(2022)]%
        {ma2022deep}
\bibfield{author}{\bibinfo{person}{Rongrong Ma}, \bibinfo{person}{Guansong
  Pang}, \bibinfo{person}{Ling Chen}, {and} \bibinfo{person}{Anton van~den
  Hengel}.} \bibinfo{year}{2022}\natexlab{}.
\newblock \showarticletitle{Deep graph-level anomaly detection by glocal
  knowledge distillation}. In \bibinfo{booktitle}{\emph{Proceedings of the
  Fifteenth ACM International Conference on Web Search and Data Mining}}.
  \bibinfo{pages}{704--714}.
\newblock


\bibitem[Makanju et~al\mbox{.}(2009)]%
        {makanju2009clustering}
\bibfield{author}{\bibinfo{person}{Adetokunbo~AO Makanju},
  \bibinfo{person}{A~Nur Zincir-Heywood}, {and} \bibinfo{person}{Evangelos~E
  Milios}.} \bibinfo{year}{2009}\natexlab{}.
\newblock \showarticletitle{Clustering event logs using iterative
  partitioning}. In \bibinfo{booktitle}{\emph{Proceedings of the 15th ACM
  SIGKDD international conference on Knowledge discovery and data mining}}.
  \bibinfo{pages}{1255--1264}.
\newblock


\bibitem[Meng et~al\mbox{.}(2020)]%
        {meng2020semantic}
\bibfield{author}{\bibinfo{person}{Weibin Meng}, \bibinfo{person}{Ying Liu},
  \bibinfo{person}{Yuheng Huang}, \bibinfo{person}{Shenglin Zhang},
  \bibinfo{person}{Federico Zaiter}, \bibinfo{person}{Bingjin Chen}, {and}
  \bibinfo{person}{Dan Pei}.} \bibinfo{year}{2020}\natexlab{}.
\newblock \showarticletitle{A semantic-aware representation framework for
  online log analysis}. In \bibinfo{booktitle}{\emph{2020 29th International
  Conference on Computer Communications and Networks (ICCCN)}}. IEEE,
  \bibinfo{pages}{1--7}.
\newblock


\bibitem[Meng et~al\mbox{.}(2019)]%
        {meng2019loganomaly}
\bibfield{author}{\bibinfo{person}{Weibin Meng}, \bibinfo{person}{Ying Liu},
  \bibinfo{person}{Yichen Zhu}, \bibinfo{person}{Shenglin Zhang},
  \bibinfo{person}{Dan Pei}, \bibinfo{person}{Yuqing Liu},
  \bibinfo{person}{Yihao Chen}, \bibinfo{person}{Ruizhi Zhang},
  \bibinfo{person}{Shimin Tao}, \bibinfo{person}{Pei Sun}, {et~al\mbox{.}}}
  \bibinfo{year}{2019}\natexlab{}.
\newblock \showarticletitle{Loganomaly: Unsupervised detection of sequential
  and quantitative anomalies in unstructured logs.}. In
  \bibinfo{booktitle}{\emph{IJCAI}}, Vol.~\bibinfo{volume}{19}.
  \bibinfo{pages}{4739--4745}.
\newblock


\bibitem[Nedelkoski et~al\mbox{.}(2020)]%
        {nedelkoski2020self}
\bibfield{author}{\bibinfo{person}{Sasho Nedelkoski}, \bibinfo{person}{Jasmin
  Bogatinovski}, \bibinfo{person}{Alexander Acker}, \bibinfo{person}{Jorge
  Cardoso}, {and} \bibinfo{person}{Odej Kao}.} \bibinfo{year}{2020}\natexlab{}.
\newblock \showarticletitle{Self-attentive classification-based anomaly
  detection in unstructured logs}. In \bibinfo{booktitle}{\emph{2020 IEEE
  International Conference on Data Mining (ICDM)}}. IEEE,
  \bibinfo{pages}{1196--1201}.
\newblock


\bibitem[Nguyen et~al\mbox{.}(2016)]%
        {nguyen2016integrating}
\bibfield{author}{\bibinfo{person}{Kim~Anh Nguyen}, \bibinfo{person}{Sabine
  Schulte~im Walde}, {and} \bibinfo{person}{Ngoc~Thang Vu}.}
  \bibinfo{year}{2016}\natexlab{}.
\newblock \showarticletitle{Integrating distributional lexical contrast into
  word embeddings for antonym-synonym distinction}.
\newblock \bibinfo{journal}{\emph{arXiv preprint arXiv:1605.07766}}
  (\bibinfo{year}{2016}).
\newblock


\bibitem[Ni et~al\mbox{.}(2022)]%
        {ni2022best}
\bibfield{author}{\bibinfo{person}{Chao Ni}, \bibinfo{person}{Wei Wang},
  \bibinfo{person}{Kaiwen Yang}, \bibinfo{person}{Xin Xia},
  \bibinfo{person}{Kui Liu}, {and} \bibinfo{person}{David Lo}.}
  \bibinfo{year}{2022}\natexlab{}.
\newblock \showarticletitle{The best of both worlds: integrating semantic
  features with expert features for defect prediction and localization}. In
  \bibinfo{booktitle}{\emph{Proceedings of the 30th ACM Joint European Software
  Engineering Conference and Symposium on the Foundations of Software
  Engineering}}. \bibinfo{pages}{672--683}.
\newblock


\bibitem[Pennington et~al\mbox{.}(2014)]%
        {pennington2014glove}
\bibfield{author}{\bibinfo{person}{Jeffrey Pennington},
  \bibinfo{person}{Richard Socher}, {and} \bibinfo{person}{Christopher~D
  Manning}.} \bibinfo{year}{2014}\natexlab{}.
\newblock \showarticletitle{Glove: Global vectors for word representation}. In
  \bibinfo{booktitle}{\emph{Proceedings of the 2014 conference on empirical
  methods in natural language processing (EMNLP)}}.
  \bibinfo{pages}{1532--1543}.
\newblock


\bibitem[Ruff et~al\mbox{.}(2018)]%
        {ruff2018deep}
\bibfield{author}{\bibinfo{person}{Lukas Ruff}, \bibinfo{person}{Robert
  Vandermeulen}, \bibinfo{person}{Nico Goernitz}, \bibinfo{person}{Lucas
  Deecke}, \bibinfo{person}{Shoaib~Ahmed Siddiqui}, \bibinfo{person}{Alexander
  Binder}, \bibinfo{person}{Emmanuel M{\"u}ller}, {and} \bibinfo{person}{Marius
  Kloft}.} \bibinfo{year}{2018}\natexlab{}.
\newblock \showarticletitle{Deep one-class classification}. In
  \bibinfo{booktitle}{\emph{International conference on machine learning}}.
  PMLR, \bibinfo{pages}{4393--4402}.
\newblock


\bibitem[Salehi et~al\mbox{.}(2021)]%
        {salehi2021multiresolution}
\bibfield{author}{\bibinfo{person}{Mohammadreza Salehi},
  \bibinfo{person}{Niousha Sadjadi}, \bibinfo{person}{Soroosh Baselizadeh},
  \bibinfo{person}{Mohammad~H Rohban}, {and} \bibinfo{person}{Hamid~R Rabiee}.}
  \bibinfo{year}{2021}\natexlab{}.
\newblock \showarticletitle{Multiresolution knowledge distillation for anomaly
  detection}. In \bibinfo{booktitle}{\emph{Proceedings of the IEEE/CVF
  conference on computer vision and pattern recognition}}.
  \bibinfo{pages}{14902--14912}.
\newblock


\bibitem[Sch{\"o}lkopf et~al\mbox{.}(2001)]%
        {scholkopf2001estimating}
\bibfield{author}{\bibinfo{person}{Bernhard Sch{\"o}lkopf},
  \bibinfo{person}{John~C Platt}, \bibinfo{person}{John Shawe-Taylor},
  \bibinfo{person}{Alex~J Smola}, {and} \bibinfo{person}{Robert~C Williamson}.}
  \bibinfo{year}{2001}\natexlab{}.
\newblock \showarticletitle{Estimating the support of a high-dimensional
  distribution}.
\newblock \bibinfo{journal}{\emph{Neural computation}} \bibinfo{volume}{13},
  \bibinfo{number}{7} (\bibinfo{year}{2001}), \bibinfo{pages}{1443--1471}.
\newblock


\bibitem[Tan et~al\mbox{.}(2023)]%
        {tan2023s2gae}
\bibfield{author}{\bibinfo{person}{Qiaoyu Tan}, \bibinfo{person}{Ninghao Liu},
  \bibinfo{person}{Xiao Huang}, \bibinfo{person}{Soo-Hyun Choi},
  \bibinfo{person}{Li Li}, \bibinfo{person}{Rui Chen}, {and}
  \bibinfo{person}{Xia Hu}.} \bibinfo{year}{2023}\natexlab{}.
\newblock \showarticletitle{S2GAE: Self-Supervised Graph Autoencoders are
  Generalizable Learners with Graph Masking}. In
  \bibinfo{booktitle}{\emph{Proceedings of the Sixteenth ACM International
  Conference on Web Search and Data Mining}}. \bibinfo{pages}{787--795}.
\newblock


\bibitem[USENIX(2008)]%
        {usenix}
\bibfield{author}{\bibinfo{person}{USENIX}.} \bibinfo{year}{2008}\natexlab{}.
\newblock \bibinfo{title}{CFDR DATA}.
\newblock \bibinfo{howpublished}{GitHub}.
\newblock
\urldef\tempurl%
\url{https://www.usenix.org/cfdr-data}
\showURL{%
\tempurl}
\newblock
\shownote{Accessed: 3/3/2023}.


\bibitem[Wan et~al\mbox{.}(2021)]%
        {wan2021glad}
\bibfield{author}{\bibinfo{person}{Yi Wan}, \bibinfo{person}{Yilin Liu},
  \bibinfo{person}{Dong Wang}, {and} \bibinfo{person}{Yujin Wen}.}
  \bibinfo{year}{2021}\natexlab{}.
\newblock \showarticletitle{GLAD-PAW: Graph-Based Log Anomaly Detection by
  Position Aware Weighted Graph Attention Network}. In
  \bibinfo{booktitle}{\emph{Pacific-Asia Conference on Knowledge Discovery and
  Data Mining}}. Springer, \bibinfo{pages}{66--77}.
\newblock


\bibitem[Wang et~al\mbox{.}(2021)]%
        {wang2021multi}
\bibfield{author}{\bibinfo{person}{Zhiwei Wang}, \bibinfo{person}{Zhengzhang
  Chen}, \bibinfo{person}{Jingchao Ni}, \bibinfo{person}{Hui Liu},
  \bibinfo{person}{Haifeng Chen}, {and} \bibinfo{person}{Jiliang Tang}.}
  \bibinfo{year}{2021}\natexlab{}.
\newblock \showarticletitle{Multi-scale one-class recurrent neural networks for
  discrete event sequence anomaly detection}. In
  \bibinfo{booktitle}{\emph{Proceedings of the 27th ACM SIGKDD conference on
  knowledge discovery \& data mining}}. \bibinfo{pages}{3726--3734}.
\newblock


\bibitem[Xiao et~al\mbox{.}(2021)]%
        {xiao2021unsupervised}
\bibfield{author}{\bibinfo{person}{Qinfeng Xiao}, \bibinfo{person}{Jing Wang},
  \bibinfo{person}{Youfang Lin}, \bibinfo{person}{Wenbo Gongsa},
  \bibinfo{person}{Ganghui Hu}, \bibinfo{person}{Menggang Li}, {and}
  \bibinfo{person}{Fang Wang}.} \bibinfo{year}{2021}\natexlab{}.
\newblock \showarticletitle{Unsupervised anomaly detection with distillated
  teacher-student network ensemble}.
\newblock \bibinfo{journal}{\emph{Entropy}} \bibinfo{volume}{23},
  \bibinfo{number}{2} (\bibinfo{year}{2021}), \bibinfo{pages}{201}.
\newblock


\bibitem[Xie et~al\mbox{.}(2022a)]%
        {xie2022self}
\bibfield{author}{\bibinfo{person}{Yaochen Xie}, \bibinfo{person}{Zhao Xu},
  {and} \bibinfo{person}{Shuiwang Ji}.} \bibinfo{year}{2022}\natexlab{a}.
\newblock \showarticletitle{Self-supervised representation learning via latent
  graph prediction}. In \bibinfo{booktitle}{\emph{International Conference on
  Machine Learning}}. PMLR, \bibinfo{pages}{24460--24477}.
\newblock


\bibitem[Xie et~al\mbox{.}(2022b)]%
        {xie2022loggd}
\bibfield{author}{\bibinfo{person}{Yongzheng Xie}, \bibinfo{person}{Hongyu
  Zhang}, {and} \bibinfo{person}{Muhammad~Ali Babar}.}
  \bibinfo{year}{2022}\natexlab{b}.
\newblock \showarticletitle{LogGD: Detecting Anomalies from System Logs by
  Graph Neural Networks}.
\newblock \bibinfo{journal}{\emph{arXiv preprint arXiv:2209.07869}}
  (\bibinfo{year}{2022}).
\newblock


\bibitem[Xie et~al\mbox{.}(2021)]%
        {xie2021logdp}
\bibfield{author}{\bibinfo{person}{Yongzheng Xie}, \bibinfo{person}{Hongyu
  Zhang}, \bibinfo{person}{Bo Zhang}, \bibinfo{person}{Muhammad~Ali Babar},
  {and} \bibinfo{person}{Sha Lu}.} \bibinfo{year}{2021}\natexlab{}.
\newblock \showarticletitle{LogDP: Combining Dependency and Proximity for
  Log-Based Anomaly Detection}. In \bibinfo{booktitle}{\emph{International
  Conference on Service-Oriented Computing}}. Springer,
  \bibinfo{pages}{708--716}.
\newblock


\bibitem[Xu et~al\mbox{.}(2023)]%
        {xu2023fascinating}
\bibfield{author}{\bibinfo{person}{Hongzuo Xu}, \bibinfo{person}{Yijie Wang},
  \bibinfo{person}{Juhui Wei}, \bibinfo{person}{Songlei Jian},
  \bibinfo{person}{Yizhou Li}, {and} \bibinfo{person}{Ning Liu}.}
  \bibinfo{year}{2023}\natexlab{}.
\newblock \showarticletitle{Fascinating Supervisory Signals and Where to Find
  Them: Deep Anomaly Detection with Scale Learning}. In
  \bibinfo{booktitle}{\emph{Proceedings of the 40th International Conference on
  Machine Learning (Poster), ICML}}.
\newblock


\bibitem[Xu et~al\mbox{.}(2009a)]%
        {xu2009largescale}
\bibfield{author}{\bibinfo{person}{Wei Xu}, \bibinfo{person}{Ling Huang},
  \bibinfo{person}{Armando Fox}, \bibinfo{person}{David Patterson}, {and}
  \bibinfo{person}{Michael Jordan}.} \bibinfo{year}{2009}\natexlab{a}.
\newblock \showarticletitle{Largescale system problem detection by mining
  console logs}.
\newblock \bibinfo{journal}{\emph{Proceedings of SOSP’09}}
  (\bibinfo{year}{2009}).
\newblock


\bibitem[Xu et~al\mbox{.}(2009b)]%
        {xu2009detecting}
\bibfield{author}{\bibinfo{person}{Wei Xu}, \bibinfo{person}{Ling Huang},
  \bibinfo{person}{Armando Fox}, \bibinfo{person}{David Patterson}, {and}
  \bibinfo{person}{Michael~I Jordan}.} \bibinfo{year}{2009}\natexlab{b}.
\newblock \showarticletitle{Detecting large-scale system problems by mining
  console logs}. In \bibinfo{booktitle}{\emph{Proceedings of the ACM SIGOPS
  22nd symposium on Operating systems principles}}. \bibinfo{pages}{117--132}.
\newblock


\bibitem[Yang et~al\mbox{.}(2021)]%
        {yang2021semi}
\bibfield{author}{\bibinfo{person}{Lin Yang}, \bibinfo{person}{Junjie Chen},
  \bibinfo{person}{Zan Wang}, \bibinfo{person}{Weijing Wang},
  \bibinfo{person}{Jiajun Jiang}, \bibinfo{person}{Xuyuan Dong}, {and}
  \bibinfo{person}{Wenbin Zhang}.} \bibinfo{year}{2021}\natexlab{}.
\newblock \showarticletitle{Semi-supervised log-based anomaly detection via
  probabilistic label estimation}. In \bibinfo{booktitle}{\emph{2021 IEEE/ACM
  43rd International Conference on Software Engineering (ICSE)}}. IEEE,
  \bibinfo{pages}{1448--1460}.
\newblock


\bibitem[Yao and Shepperd(2021)]%
        {yao2021impact}
\bibfield{author}{\bibinfo{person}{Jingxiu Yao} {and} \bibinfo{person}{Martin
  Shepperd}.} \bibinfo{year}{2021}\natexlab{}.
\newblock \showarticletitle{The impact of using biased performance metrics on
  software defect prediction research}.
\newblock \bibinfo{journal}{\emph{Information and Software Technology}}
  \bibinfo{volume}{139} (\bibinfo{year}{2021}), \bibinfo{pages}{106664}.
\newblock


\bibitem[Zhang et~al\mbox{.}(2020)]%
        {zhang2020anomaly}
\bibfield{author}{\bibinfo{person}{Bo Zhang}, \bibinfo{person}{Hongyu Zhang},
  \bibinfo{person}{Pablo Moscato}, {and} \bibinfo{person}{Aozhong Zhang}.}
  \bibinfo{year}{2020}\natexlab{}.
\newblock \showarticletitle{Anomaly detection via mining numerical workflow
  relations from logs}. In \bibinfo{booktitle}{\emph{2020 International
  Symposium on Reliable Distributed Systems (SRDS)}}. IEEE,
  \bibinfo{pages}{195--204}.
\newblock


\bibitem[Zhang et~al\mbox{.}(2022)]%
        {zhang2022cat}
\bibfield{author}{\bibinfo{person}{Shengming Zhang}, \bibinfo{person}{Yanchi
  Liu}, \bibinfo{person}{Xuchao Zhang}, \bibinfo{person}{Wei Cheng},
  \bibinfo{person}{Haifeng Chen}, {and} \bibinfo{person}{Hui Xiong}.}
  \bibinfo{year}{2022}\natexlab{}.
\newblock \showarticletitle{Cat: Beyond efficient transformer for content-aware
  anomaly detection in event sequences}. In
  \bibinfo{booktitle}{\emph{Proceedings of the 28th ACM SIGKDD Conference on
  Knowledge Discovery and Data Mining}}. \bibinfo{pages}{4541--4550}.
\newblock


\bibitem[Zhang et~al\mbox{.}(2019)]%
        {zhang2019robust}
\bibfield{author}{\bibinfo{person}{Xu Zhang}, \bibinfo{person}{Yong Xu},
  \bibinfo{person}{Qingwei Lin}, \bibinfo{person}{Bo Qiao},
  \bibinfo{person}{Hongyu Zhang}, \bibinfo{person}{Yingnong Dang},
  \bibinfo{person}{Chunyu Xie}, \bibinfo{person}{Xinsheng Yang},
  \bibinfo{person}{Qian Cheng}, \bibinfo{person}{Ze Li}, {et~al\mbox{.}}}
  \bibinfo{year}{2019}\natexlab{}.
\newblock \showarticletitle{Robust log-based anomaly detection on unstable log
  data}. In \bibinfo{booktitle}{\emph{Proceedings of the 2019 27th ACM Joint
  Meeting on European Software Engineering Conference and Symposium on the
  Foundations of Software Engineering}}. \bibinfo{pages}{807--817}.
\newblock


\bibitem[Zhu et~al\mbox{.}(2023a)]%
        {loghub_website}
\bibfield{author}{\bibinfo{person}{Jieming Zhu}, \bibinfo{person}{Shilin He},
  \bibinfo{person}{Pinjia He}, \bibinfo{person}{Jinyang Liu}, {and}
  \bibinfo{person}{Michael~R. Lyu}.} \bibinfo{year}{2023}\natexlab{a}.
\newblock \bibinfo{howpublished}{GitHub}.
\newblock
\urldef\tempurl%
\url{https://github.com/logpai/loghub}
\showURL{%
\tempurl}


\bibitem[Zhu et~al\mbox{.}(2023b)]%
        {loghub}
\bibfield{author}{\bibinfo{person}{Jieming Zhu}, \bibinfo{person}{Shilin He},
  \bibinfo{person}{Pinjia He}, \bibinfo{person}{Jinyang Liu}, {and}
  \bibinfo{person}{Michael~R. Lyu}.} \bibinfo{year}{2023}\natexlab{b}.
\newblock \showarticletitle{Loghub: {A} Large Collection of System Log Datasets
  for AI-driven Log Analytics}. In \bibinfo{booktitle}{\emph{IEEE International
  Symposium on Software Reliability Engineering (ISSRE)}}.
\newblock


\bibitem[Zhu et~al\mbox{.}(2019)]%
        {zhu2019tools}
\bibfield{author}{\bibinfo{person}{Jieming Zhu}, \bibinfo{person}{Shilin He},
  \bibinfo{person}{Jinyang Liu}, \bibinfo{person}{Pinjia He},
  \bibinfo{person}{Qi Xie}, \bibinfo{person}{Zibin Zheng}, {and}
  \bibinfo{person}{Michael~R Lyu}.} \bibinfo{year}{2019}\natexlab{}.
\newblock \showarticletitle{Tools and benchmarks for automated log parsing}. In
  \bibinfo{booktitle}{\emph{2019 IEEE/ACM 41st International Conference on
  Software Engineering: Software Engineering in Practice (ICSE-SEIP)}}. IEEE,
  \bibinfo{pages}{121--130}.
\newblock


\end{thebibliography}

\appendix

\end{document}